\documentclass[aip,reprint,amsmath]{revtex4-1}

\usepackage{multirow}
\usepackage{graphicx}
\usepackage{subfigure}
\usepackage{color}

\begin{document}


\title[]{A quadratic Reynolds stress development for the turbulent Kolmogorov flow} 



\author{Wenwei Wu}
\email{wenwei\_wu@sjtu.edu.cn}
\affiliation{Univ. Lille, CNRS, ULCO, Laboratory of Oceanology and Geosciences, UMR LOG 8187, Wimereux, France}
\affiliation{UM-SJTU Joint Institute, Shanghai JiaoTong University 200240, Shanghai, People's Republic of China}
\author{Fran\c{c}ois G. Schmitt }
\email{francois.schmitt@cnrs.fr}
\affiliation{Univ. Lille, CNRS, ULCO, Laboratory of Oceanology and Geosciences, UMR LOG 8187, Wimereux, France}
\author{Enrico Calzavarini }
\affiliation{Univ. Lille, Unit\'e de M\'ecanique de Lille - J. Boussinesq (UML) ULR 7512, F59000 Lille, France}
\author{Lipo Wang }
\affiliation{UM-SJTU Joint Institute, Shanghai JiaoTong University 200240, Shanghai, People's Republic of China}


\date{\today}

\begin{abstract}
We study the three-dimensional turbulent Kolmogorov flow, i.e. the Navier-Stokes equations forced by a
single-low-wave-number sinusoidal force in a periodic domain, by means of direct numerical simulations.
This classical model system is a realization of anisotropic and non-homogeneous hydrodynamic turbulence.
Boussinesq's eddy viscosity linear relation is checked and found
to be approximately valid over half of the system volume.
A more general quadratic Reynolds stress development is proposed
and its parameters estimated at varying the Taylor scale-based Reynolds number in the flow up to the value 200.
The case of a forcing with a different shape, here chosen Gaussian, is considered and the differences with the sinusoidal forcing are emphasized.
\end{abstract}


\maketitle 


\section{Introduction}\label{intro}
In the late 1950s, A.N. Kolmogorov proposed to study the stability properties of an incompressible flow described by the Navier-Stokes equation forced by a sinusoidal shear force in a periodic domain.
An answer was put forward soon after \cite{Meshalkin1961}, indicating the existence of a critical Reynolds number $R = \sqrt{2}$,
confirmed also later in \cite{Green1974} (we will provide below the
definition of Reynolds number which was used in these works).
Such a model system, since then dubbed Kolmogorov flow (KF), is
not straightforward to be realized in experiments.
However, an important work published in the Russian literature \cite{Bondarenko1979}
and discussed by A. N. Obukhov \cite{Obukhov1983} describes
an experiment using a thin layer of electrolyte in an external force field capable to generate an analogous flow, for which the stability properties as well as the transition to the turbulent state were studied. The results of this experiment, supported by a previous
theoretical work \cite{Kliatskin1972}, have been taken as an evidence that there exists in the KF a succession of instabilities with increasing Reynolds numbers, until reaching a fully turbulent state for
Reynolds numbers of the order of 1000 \cite{Bondarenko1979,Obukhov1983}.
One the other hand, since the advent of computers and in particular since the introduction of the Fast Fourier Transform (FFT) algorithm the Kolmogorov flow has become amenable to be explored via numerical simulations, even in high-Reynolds number conditions \cite{Borue1996}.
This makes the turbulent Kolmogorov flow (TKF) possibly the simplest and most accessible prototype of open flow, i.e. a flow without boundaries, which is at the same time statistically stationary, anisotropic, and non-homogeneous (along one direction) \cite{Borue1996,Shebalin1997,Biferale2001,Balmforth2002,Boffetta2005,Musacchio2014}.
As discussed by Musacchio and Boffetta \cite{Musacchio2014}, the TKF can be considered, to some respect,  as a turbulent
channel flow (i.e. a pressure-driven parallel flow) without boundaries.
In recent years, the TKF has been mainly studied theoretically and numerically.
The large-scale forcing was originally proposed as a sinusoidal force, which, however, is not a requirement, and rather constitutes a convenient simplification for the theoretical analysis and for numerical implementations (which are mostly based on FFT). Other
shapes for the large-scale forcing could be imagined as well, see e.g. \cite{Rollin2011}.
Many numerical studies devoted to KF and TKF have adopted a two-dimensional
configuration \cite{She1987,Berti2010,Chandler2013,Lucas2014,Lucas2015}, because of its reduced computational cost. It is however known that 2D turbulence differs from the 3D one due to the existence of an inverse energy cascade.
For this reason in this work we prefer to focus on the more realistic
three-dimensional case, i.e. fully resolved Navier-Stokes incompressible turbulence forced along
the $x$
direction by a large-scale sinusoidal force depending solely on the $z$ coordinate.
As observed by \citet{Borue1996}, such flow is a convenient test ground for transport models. Such a consideration motivates the present study.\\
The structure of this article is as follows: After a section introducing the present notations and numerical 
implementation, the theoretical framework of linear and nonlinear closure equation for the Reynolds stress tensor is presented and adapted to the TKF model system.
First, we consider the classical turbulence closure based on
eddy-viscosity Boussinesq's approach, where the traceless stress tensor is assumed to be proportional to the mean strain-rate tensor. Such expression is at the basis of
many turbulence models including $k$-$\epsilon$, $k$-$\omega$ and all eddy viscosity transport models \cite{Pope2000}. We show the range of applicability and limitations of this assumption in the context of TKF.
Second, a nonlinear quadratic Reynolds stress development, that makes use of tensor invariants is directly tested on TKF. This is performed along the lines of previous
direct test done for channel flows \cite{Schmitt2007a,Schmitt2007b},
and for various Reynolds numbers. 
We also compare the different terms of the kinetic energy equation. In a following section a model flow system with a different forcing, non-sinusoidal shape, is considered and its differences with the original KF forcing are considered. The last section is devoted to a discussion of the main findings and
conclusions.

\section{The Kolmogorov flow model system}\label{DNS}

\subsection{Equations of motion and numerical implementations}
The governing equations for velocity field $\mathbf{u}(\mathbf{x},t)$ are the incompressible
Navier-Stokes equations,
\begin{equation}
   \frac{\partial \mathbf{u}}{\partial t} +(\mathbf{u}\cdot\nabla)\mathbf{u}
   = -\frac{1}{\rho}\nabla p + \nu \Delta\mathbf{u}+\mathbf{f},
\label{NS}
\end{equation}
where
$p$ is the hydrodynamic pressure,
$\rho$ the fluid density and $\nu$ is the kinematic viscosity.
This flow is sustained by a constant in time and spatially dependent force $\mathbf{f}$
of the form:
\begin{equation}\label{forcing}
\mathbf{f}=A \sin \left( 2\pi \frac{z}{H} \right)\mathbf{e}_x,
\end{equation}
where $A$ is a constant, $H$ is the length of the side of the cubic domain, chosen here as the characteristic length scale.
Such a force, directed along the $x$ direction and depending only on the $z$ coordinate, makes the turbulent flow statistically anisotropic and non-homogeneous along the $z$ direction (but statistically homogeneous in the $x$-$y$ planes).
It is convenient to introduce the following reference scales for velocity and time:
\begin{eqnarray}\label{carscale}
  U_0 &=& \left( AH \right)^{1/2} ;\\
  \label{carscale2}
  T_0 &=& \frac{H}{ U_0}=\left( \frac{H}{ A} \right)^{1/2}.
\end{eqnarray}
From this, one can construct the Reynolds number as
\begin{equation}
    Re = \frac{U_0 H}{ \nu},
\end{equation}
which thus becomes the only dimensionless control parameter in the system.
Let us mention that in stability analyses \cite{Kliatskin1972,Green1974} another
Reynolds number is used, which in the present notation reads
$R=AH^3/(\nu^2(2\pi)^3)$. It thus yields $R=Re^2/((2\pi)^3)$,
where the $2 \pi$ factor originates from a slightly different choice of the reference length: $H$ as the characteristic length for a sine wave $\sin(2 \pi z/H)$ in the present case,
 $L$ for a sine wave written $\sin(z/L)$ in \cite{Kliatskin1972}.
The stability criterion $R> \sqrt{2}$ becomes $Re> 2^{1/4}(2\pi)^{3/2}\simeq 18.7$.
In the following we will also use the Reynolds number based on Taylor microscale, $Re_\lambda= \lambda u'/ \nu$, where $\lambda=u'\sqrt{15\nu/\epsilon}$, $\nu$ is the kinematic viscosity, $u'=\frac{1}{3} \sqrt{\overline{ \mathbf{u}'^2}}$ is the global root-mean square of single component velocity, $\epsilon = \frac{\nu}{2} \overline{\sum_i \sum_j (\partial_i u_j + \partial_j u_i)^2 }$ is the global energy dissipation rate, and the overbar $\overline{\cdot \cdot \cdot}$ denotes the global average (in time and all over the spatial domain). The latter number is more convenient to quantify the degree of turbulence realized in the system. In the rest of this article, all the reported quantities are
dimensionless with reference to the units defined in (\ref{carscale}) and (\ref{carscale2}).

The Kolmogorov flow model system is numerically simulated in a cubic tri-periodic domain. The dynamical equations \eqref{NS} are solved numerically by means of a pseudo-spectral code using a smooth dealiasing technique \citep{Hou2007} for the treatment of non-linear terms in the equations. Instead of the sudden cut-off used in the conventional $2/3$ rule approach, a filter of the high wave number modes with a relatively smooth filtering function is performed for a the smooth dealiasing, which is capable to reduce numerical high frequency instabilities.
The spatial resolution is chosen such that the condition $|\vec{k}|_{max}\cdot\eta >1$, where $\eta$ is the global Kolmogorov scale (which was checked to be nearly independent of the $z$ coordinate), $|\vec{k}|_{max}\approx0.41N$ is the maximum  wave number amplitude kept by the dealiasing procedure, is always satisfied. The time-marching scheme adopts a third order Runge-Kutta method.
The global non-dimensional values of the key parameters for the simulations are reported in table~\ref{parameters}.
Two criteria have been proposed for the convergence of Kolmogorov flow simulations \cite{Sarris2007}: first the mean energy injection should be equal (within numerically accuracy) to the total dissipation, and second the left-hand-side and right-hand-side of equation \eqref{Stress} must be equal.
It has been checked here that these two criteria are satisfied in our simulations (the right-hand-side of equation \eqref{Stress} is shown in figure \ref{Fig2}(b)). The total integration time is chosen in such a way to have comparable datasets for each run and to ensure the statistical convergence of the measurements (see Table \ref{parameters}).

\begin{table*}
\caption{\label{parameters}Global key parameters in each simulation (all provided in the dimensionless units defined in eq.\eqref{carscale} and \eqref{carscale2}). The columns from left to right report respectively: the Run number; the Reynolds numbers $Re=\frac{HU_0}{ \nu}$, $Re_\lambda=\frac{\lambda u'}{\nu}$ the Taylor-scale based number, $Re^*= \frac{U H}{2 \pi \nu} = \frac{H\kappa U_0}{2\pi\nu}$ (same as in Ref \cite{Musacchio2014}); the kinematic viscosity $\nu$; $\epsilon = \frac{\nu}{2} \overline{\sum_i \sum_j (\partial_i u_j + \partial_j u_i)^2 }$ is the global energy dissipation rate, $\overline\cdot$ denotes the global average (in time and all over the spatial domain).  $N^3$ is the grid size; $\eta=(\nu^3/\epsilon)^{1/4}$ is the global Kolmogorov scale; $|\vec{k}|_{max}\cdot\eta$ is the spatial resolution condition, where  $|\vec{k}|_{max}\approx0.41N$ is the maximum  wave number amplitude kept by the dealiasing procedure; $T_{total}$ is the total simulation time and $T_l$ is the large eddy turnover time \citep{Pope2000}, which thus implies that $T_{total}/T_l$ denotes the number of large eddy turnover time spanned by the simulation in statistically steady conditions; $\Delta t$ is the numerical time step; $\nu_T$ is the turbulent viscosity calculated according to Eq. \eqref{nu_T}.}
\begin{ruledtabular}
\begin{tabular}{cccccccccccc}
 No.& $Re$ & $Re_{\lambda}$ & $Re^*$ & $\nu$ & $\epsilon$ & $\eta$ & $N^3$ & $|\vec{k}|_{max}\cdot\eta$ & $T_{total}/T_l$ & $\Delta t$ & $\nu_T$\\
 \hline
1 & 787.5 & 38.7 & 126.24 & 0.0013 & 0.51 & 0.008 & $ 128 ^3$ & 2.64 & 462.5 & 0.0015 &  0.0239 \\
2 & 984.4 & 43.5 & 160.37 & 0.001 & 0.52 & 0.0067 & $ 128 ^3$ & 2.22 & 445.0 & 0.0014 & 0.0237 \\
3 & 1211.5 & 49.3 & 197.87 & 0.00083 & 0.52 & 0.0058 & $ 128 ^3$ & 1.90 & 427.1 & 0.0014 & 0.0239 \\
4 & 1575.0 & 57.4 & 261.36 & 0.00063 & 0.52 & 0.0047 & $ 128 ^3$ & 1.56 & 403.5 & 0.0013 & 0.0236 \\
5 & 2099.9 & 66.9 & 358.34 & 0.00048 & 0.54 & 0.0038 & $ 128 ^3$ & 1.25 & 761.2 & 0.0013 & 0.0231 \\
6 & 3149.9 & 83.9 & 565.77 & 0.00032 & 0.57 & 0.0028 & $ 256 ^3$ & 1.82 & 259.4 & 0.00058 & 0.0221 \\
7 & 6299.8 & 123.4 & 1132.53 & 0.00016 & 0.56 & 0.0016 & $ 256 ^3$ & 1.08 & 386.0 & 0.00054 & 0.0223 \\
8 & 15749.6 & 198.3 & 2817.24 & 6.3e-05 & 0.56 & 0.00083 & $ 512 ^3$ & 1.09 & 70.2 & 0.00025 & 0.0225 \\
\end{tabular}
\end{ruledtabular}
\end{table*}

\subsection{Reynolds decomposition and velocity moments}
Let us consider a Reynolds decomposition of the velocity into mean and fluctuating quantities $\mathbf{u} =\langle \mathbf{u}\rangle +\mathbf{u}'$ ($\langle \cdot\rangle$ denotes the average over time and spatially along $x$ and $y$ directions) and note the three cartesian components of the velocity $\mathbf{u}=(u,v,w)$. Because of the periodicity in $x$ and $y$ directions, the derivatives with respect to $x$ and $y$ of mean quantities are $0$.
By taking the average $\langle \cdot\rangle$ of the Navier-Stokes equations, one obtains the following relations:
\begin{equation}
 \begin{aligned}
  - \partial_z \tau &= \frac{1}{Re}  \partial_{z}^{2} U(z) +\sin(2\pi z),   \\
   \partial_z\langle w'^2\rangle &= -\partial_z \langle p\rangle,
\label{eq_flow_mean}
\end{aligned}
\end{equation}
where $U= \langle \mathbf{u}\rangle= \langle u\rangle$ and the shear stress is
$\tau = -\langle u'w'\rangle$. Using the first line of this relation, one can verify that for laminar flows when $\tau = 0$, the mean velocity profile is sinusoidal while the pressure field is constant \cite{Meshalkin1961}.

\begin{figure}
\includegraphics[width=0.7\linewidth]{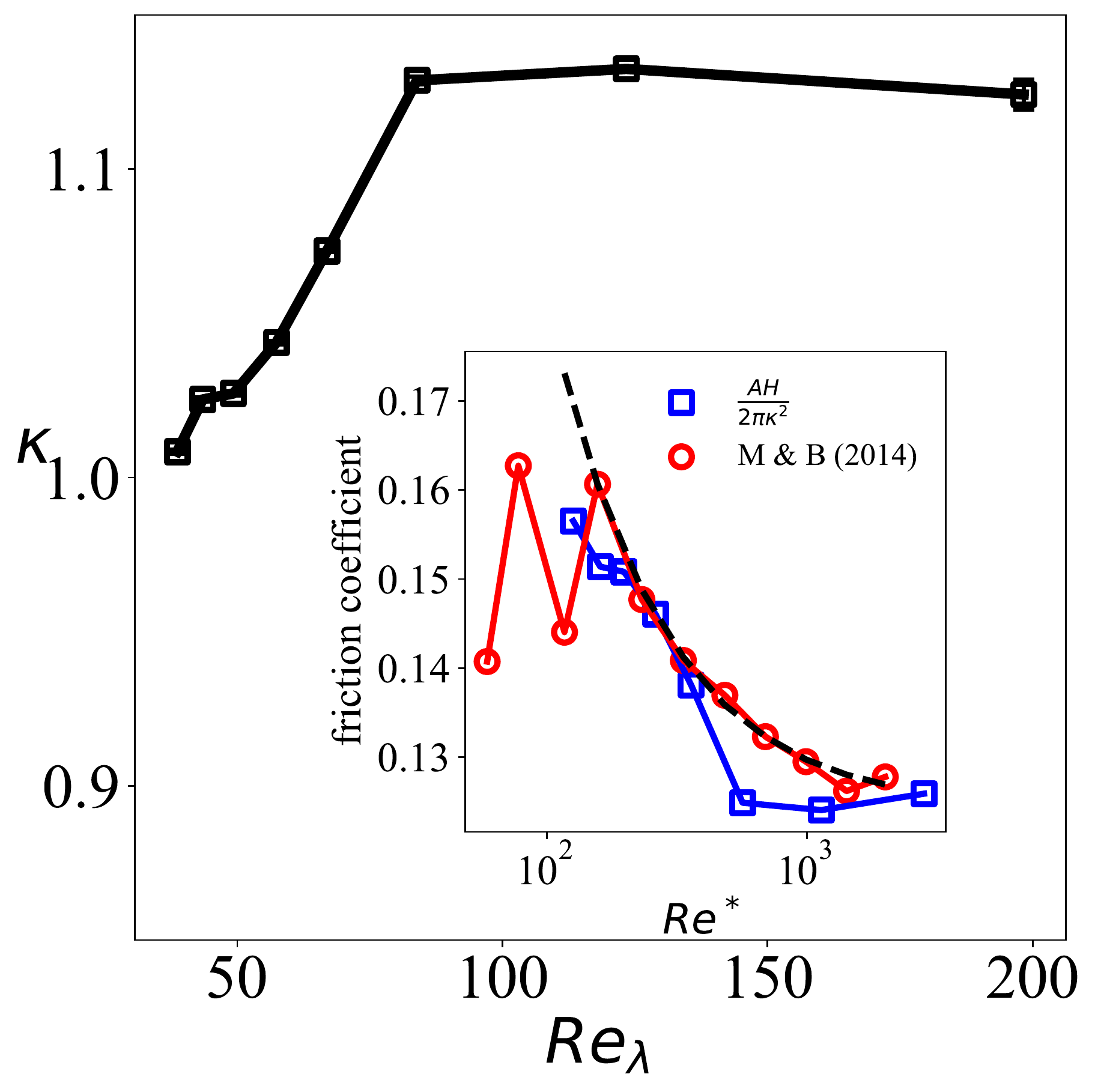}\\
\includegraphics[width=0.7\linewidth]{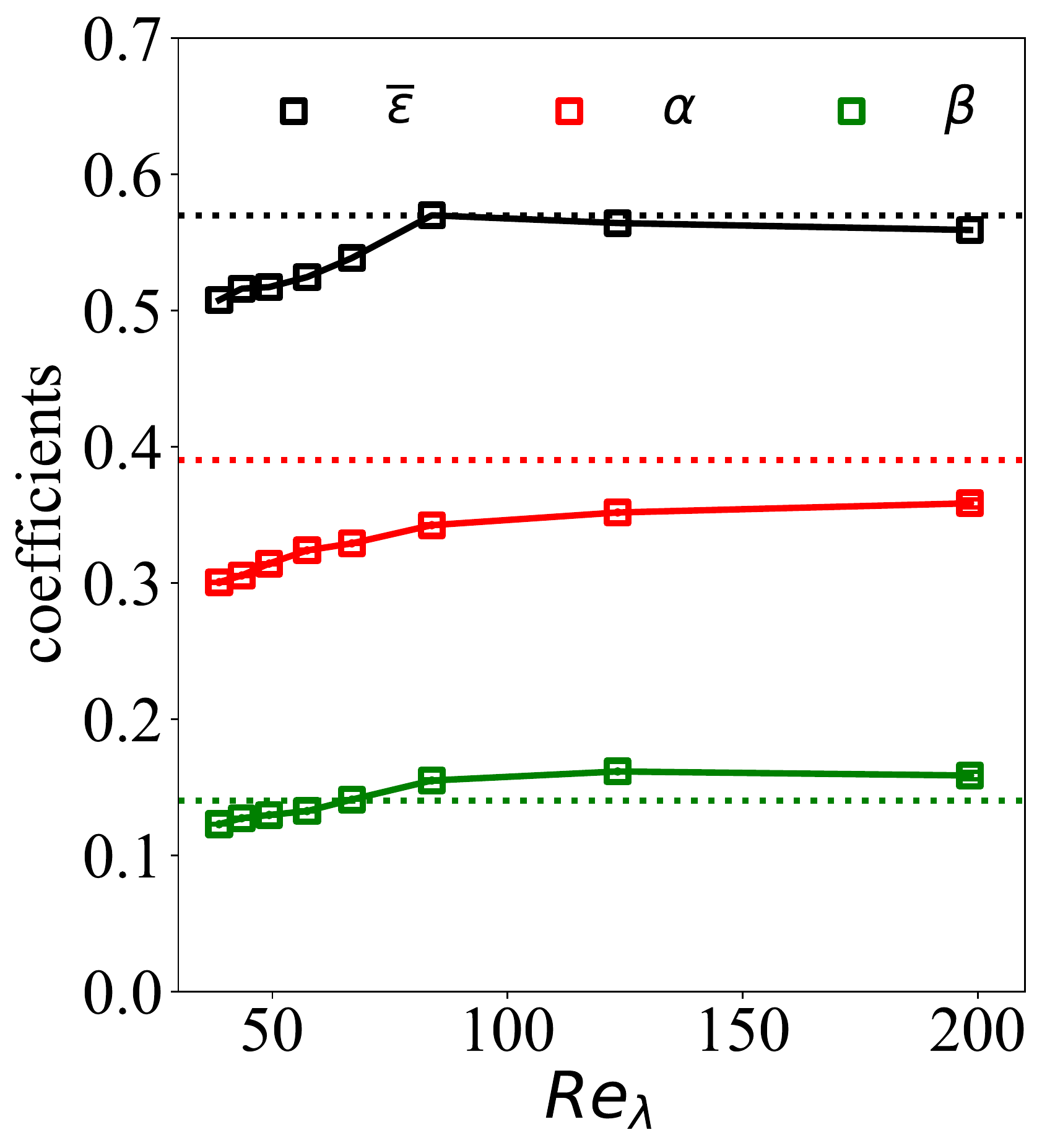}
\caption{
(a) Amplitude of  mean velocity profile, $\kappa$ in Eq. \ref{Uturb}, as function of $Re_{\lambda}$. The inset panel shows the dependence of the friction coefficient ($f$) on $Re^*$ (blue), in comparison with the result obtained by \citet{Musacchio2014} (red). The dashed black line in the inset panel shows the curve of $f=0.124+5.75/Re^*$, which fits the red dots in the large $Re^*$ range \citet{Musacchio2014}. (b) The global dissipation rate ($\overline\epsilon$) and the coefficients obtained by fitting the profiles of kinetic energy, $\alpha$ and $\beta$ in Eq. \ref{K}, as function of $Re_{\lambda}$. The horizontal dotted red and green lines represent the values $\alpha=0.391$ and $\beta=0.138$, respectively, obtained by \citet{Borue1996}. The horizontal black dotted line represents the global dissipation rate  (expressed in the dimensionless form used in this work) reported by \citet{Musacchio2014} for $Re^*=2000$.
}
\label{Kappa_alpha_beta_Vs_Re}
\end{figure}

For turbulent flows, it is well-known that the mean velocity profile is also sinusoidal \cite{Borue1996}. However, this is a numerical result which has so far,
to our knowledge, no direct analytical  explanation.
We obtain the following
$z$-dependence for $U$:
\begin{equation}
U(z)= \kappa   \sin (2 \pi z).
\label{Uturb}
\end{equation}
where $\kappa$ is a coefficient whose numerical estimation, at varying the Reynolds number, is plotted in figure \ref{Kappa_alpha_beta_Vs_Re}(b). The maximum value of the mean turbulent velocity is of the order of the characteristic velocity built using the forcing values, since we obtain values of $\kappa$ between 1.01 and 1.12, increasing with the
Reynolds number (figure \ref{Kappa_alpha_beta_Vs_Re}(a) and Table \ref{alpha_beta_1_2_3_value}). The values of $\kappa$ found here are compatible with the value of $\kappa=1.1$ 
reported by \citet{Borue1996} (however, this work makes use of hyperviscosity of the 8th order and the value of the Reynolds
number is not provided).
In the work by \citet{Musacchio2014}, the dependence of the friction coefficient $f$ (written as $f = \frac{AH}{2\pi\kappa^2}$ in the present notation), on the Reynolds number based on the forcing scale and mean velocity (denoted here $Re^*$, which writes $Re^*=\frac{\kappa}{2 \pi} Re$ in our notation) was investigated. Such dependence is also plotted in the inset of figure \ref{Kappa_alpha_beta_Vs_Re}(a): the friction coefficient values obtained in our simulations are comparable with those reported by \citet{Musacchio2014} in the same range of Reynolds numbers.

\begin{figure}
\includegraphics[width=0.95\linewidth]{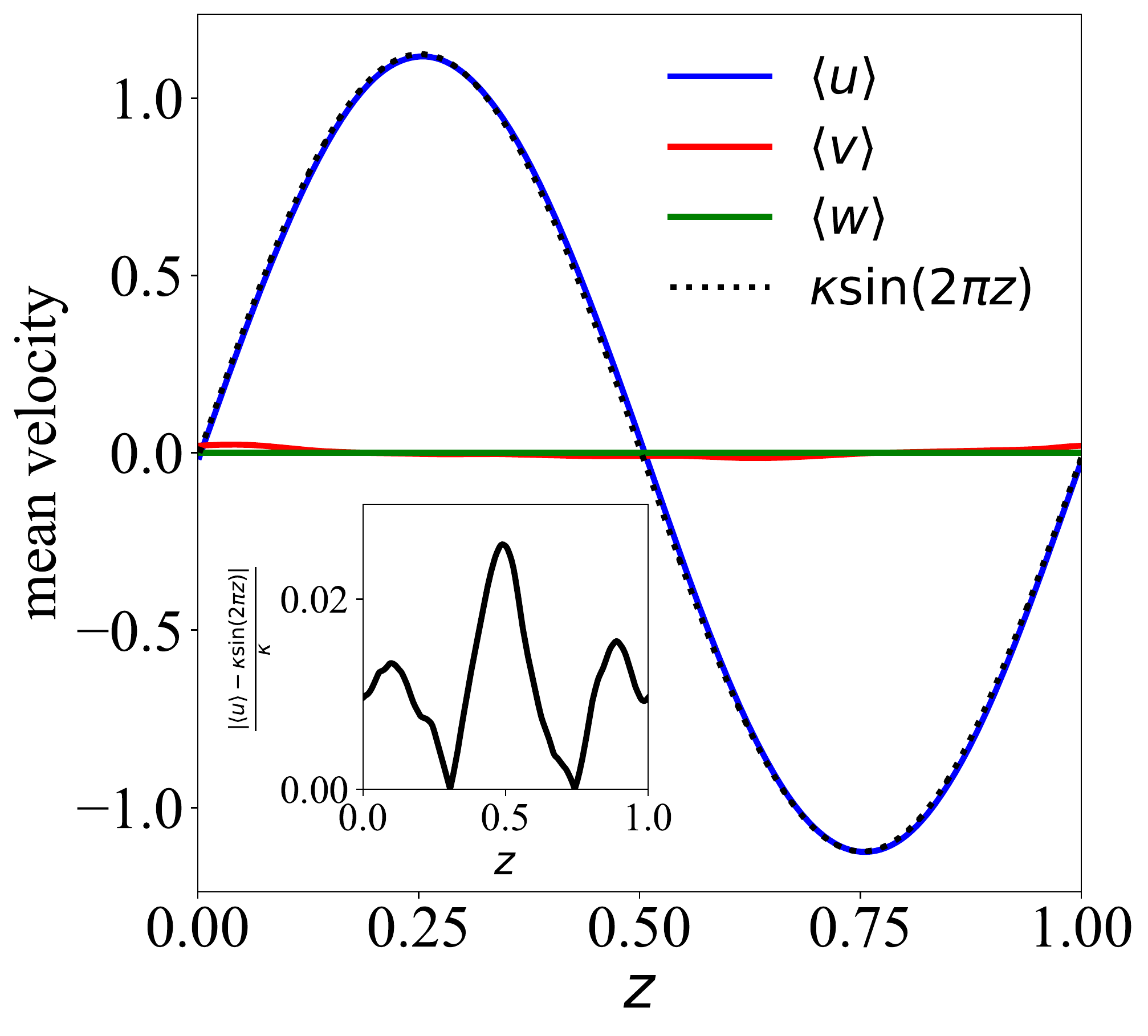}\\
\includegraphics[width=0.95\linewidth]{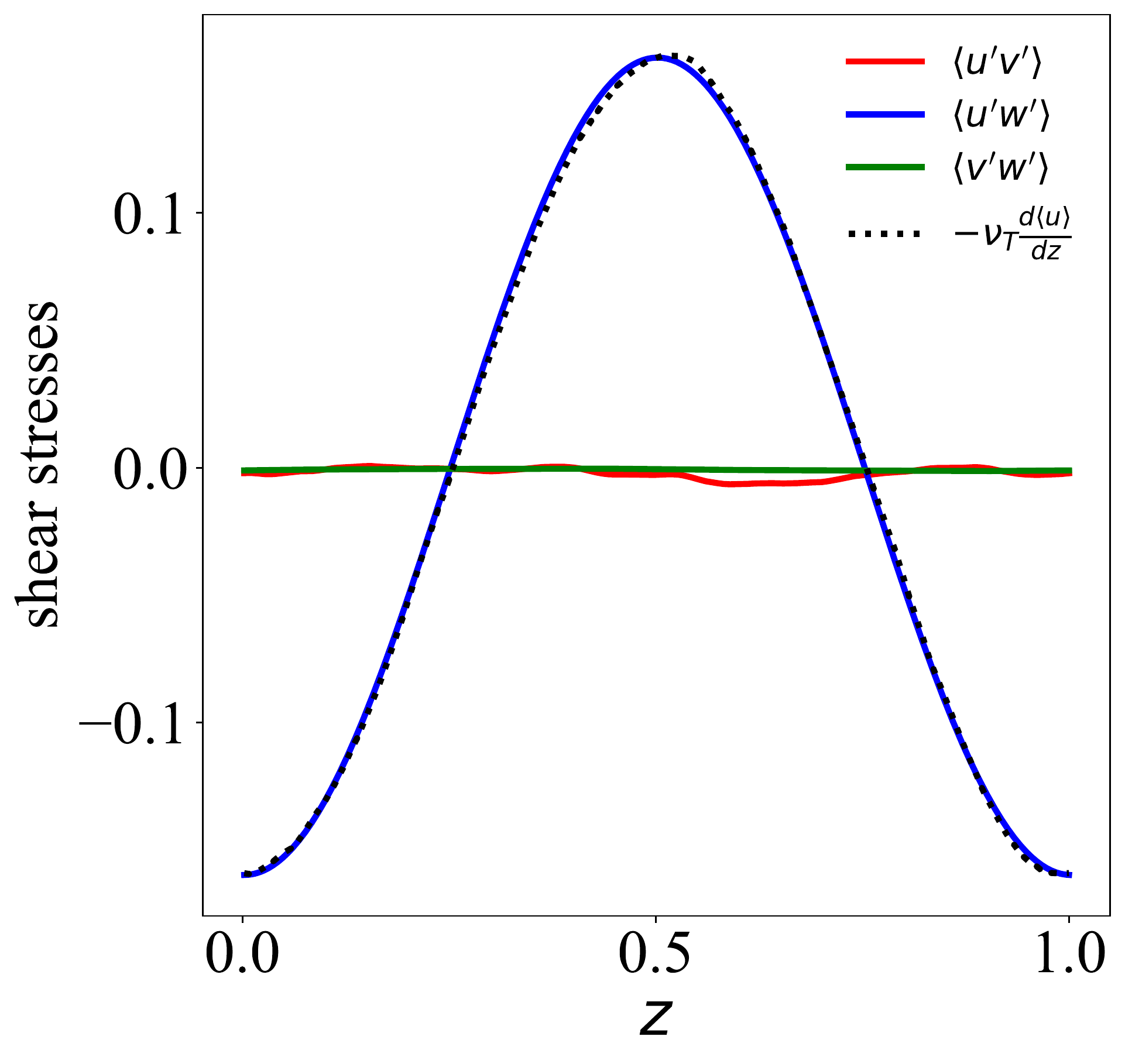}
\caption{(a) The adimensional mean quantities of each component of the velocity of Run 8. The only non-zero term is $\langle u\rangle$, having a maximum value of $\kappa $, where $\kappa = 1.12$. The black dotted line shows the curve of $\kappa\sin(2\pi z)$.
The inset plot shows the deviation of $\langle u\rangle$ from $\kappa\sin(2\pi z)$  estimated by $\frac{|\langle u\rangle-\kappa\sin(2\pi z)|}{\kappa}$.
(b) The different adimensional shear stress terms of Run 8. The only non-zero term is $\langle u'w'\rangle$, whose $z$ dependence is given by relation (\ref{Stress}). The black dotted line shows the function $-\nu_T\frac{d\langle u\rangle}{dz}$, where $\nu_T =0.023$ is the turbulent viscosity (Eq. \eqref{nu_T}) for the Run 8.
}
\label{Fig2}
\end{figure}

\begin{figure}
\centerline{
\subfigure[$Re_{\lambda}=38.7$]{\includegraphics[width=0.45\linewidth]{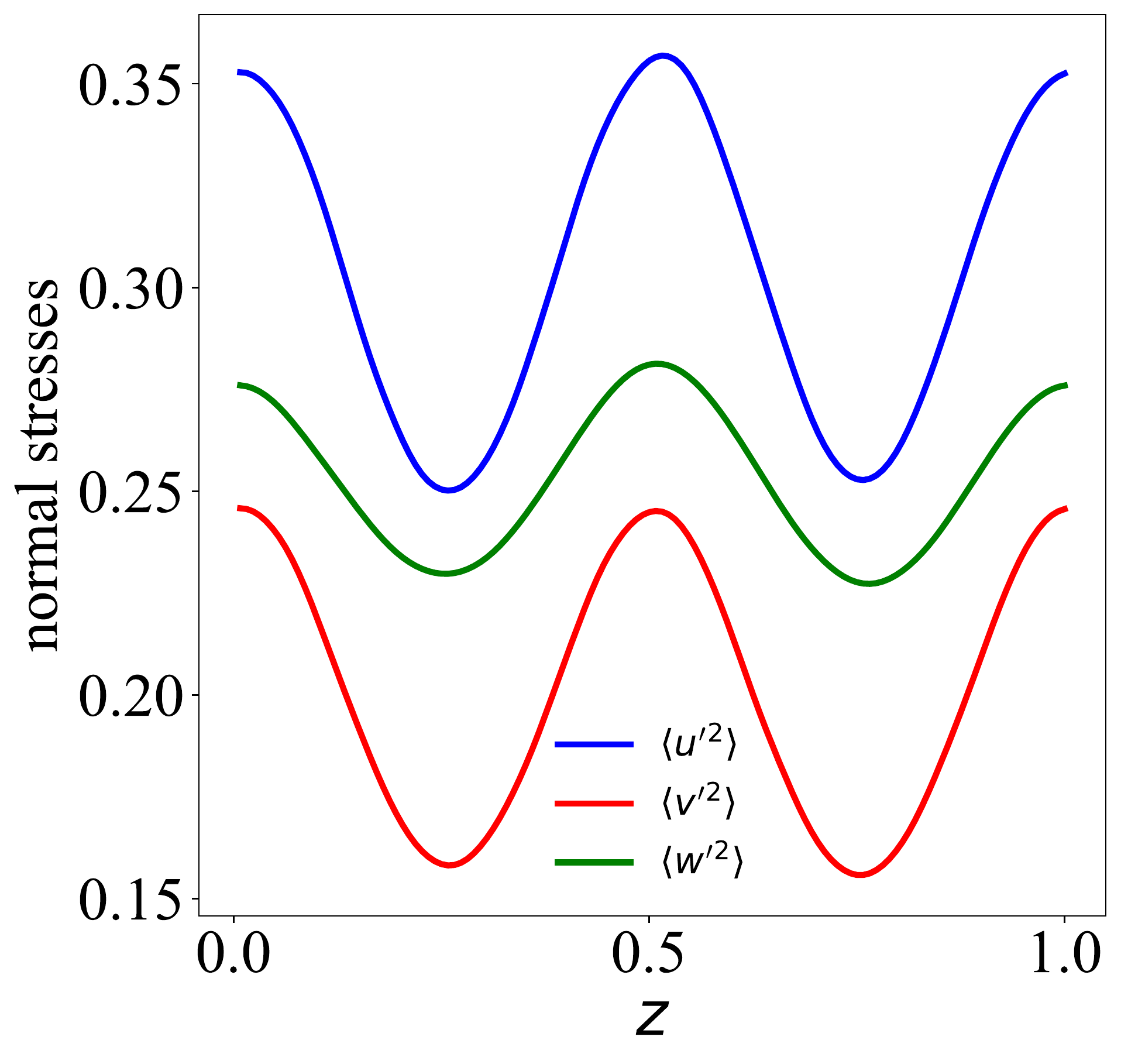}}
\subfigure[$Re_{\lambda}=49.3$]{\includegraphics[width=0.45\linewidth]{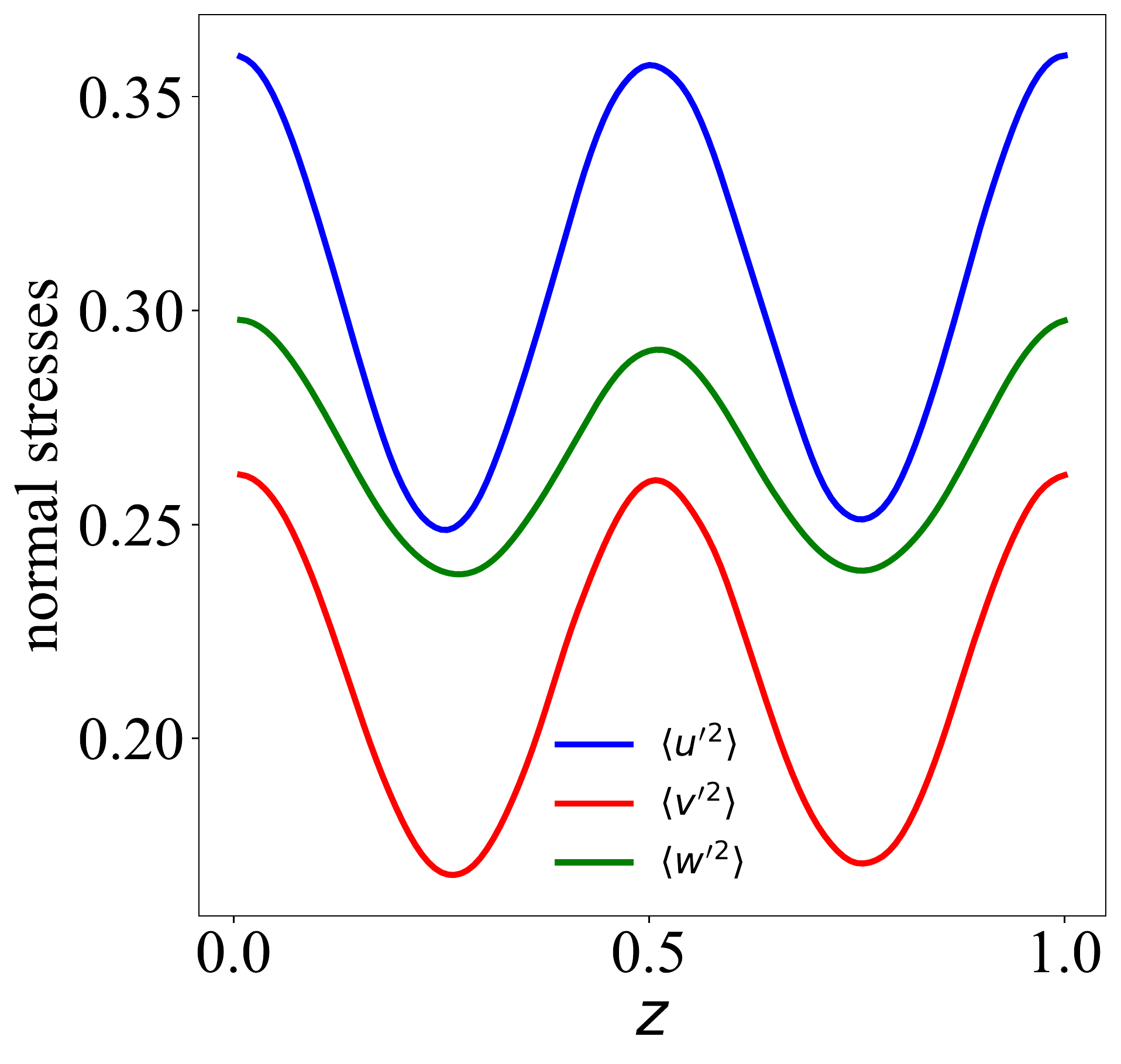}}
}
\centerline{
\subfigure[$Re_{\lambda}=66.9$]{\includegraphics[width=0.45\linewidth]{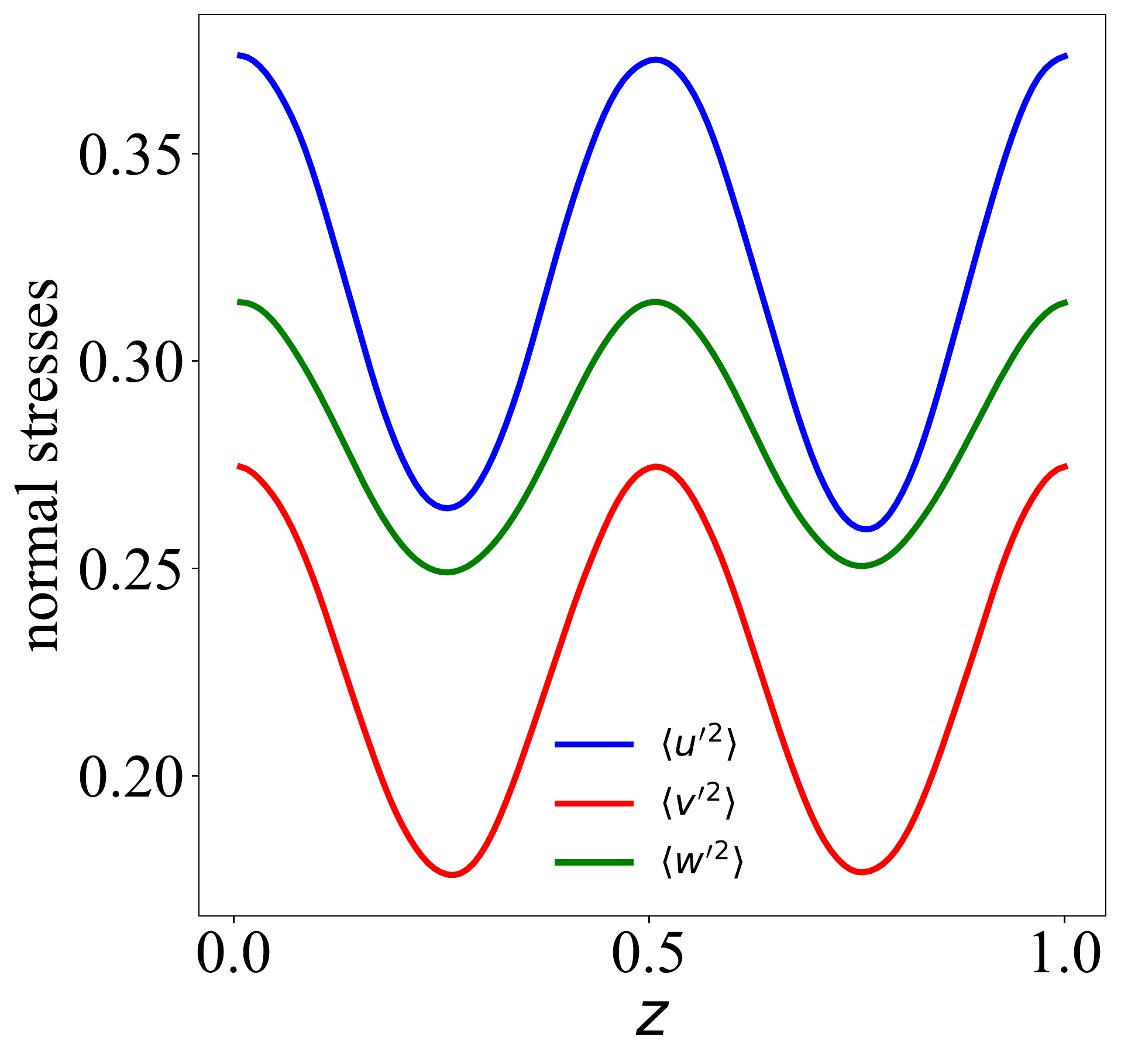}}
\subfigure[$Re_{\lambda}=123.4$]{\includegraphics[width=0.45\linewidth]{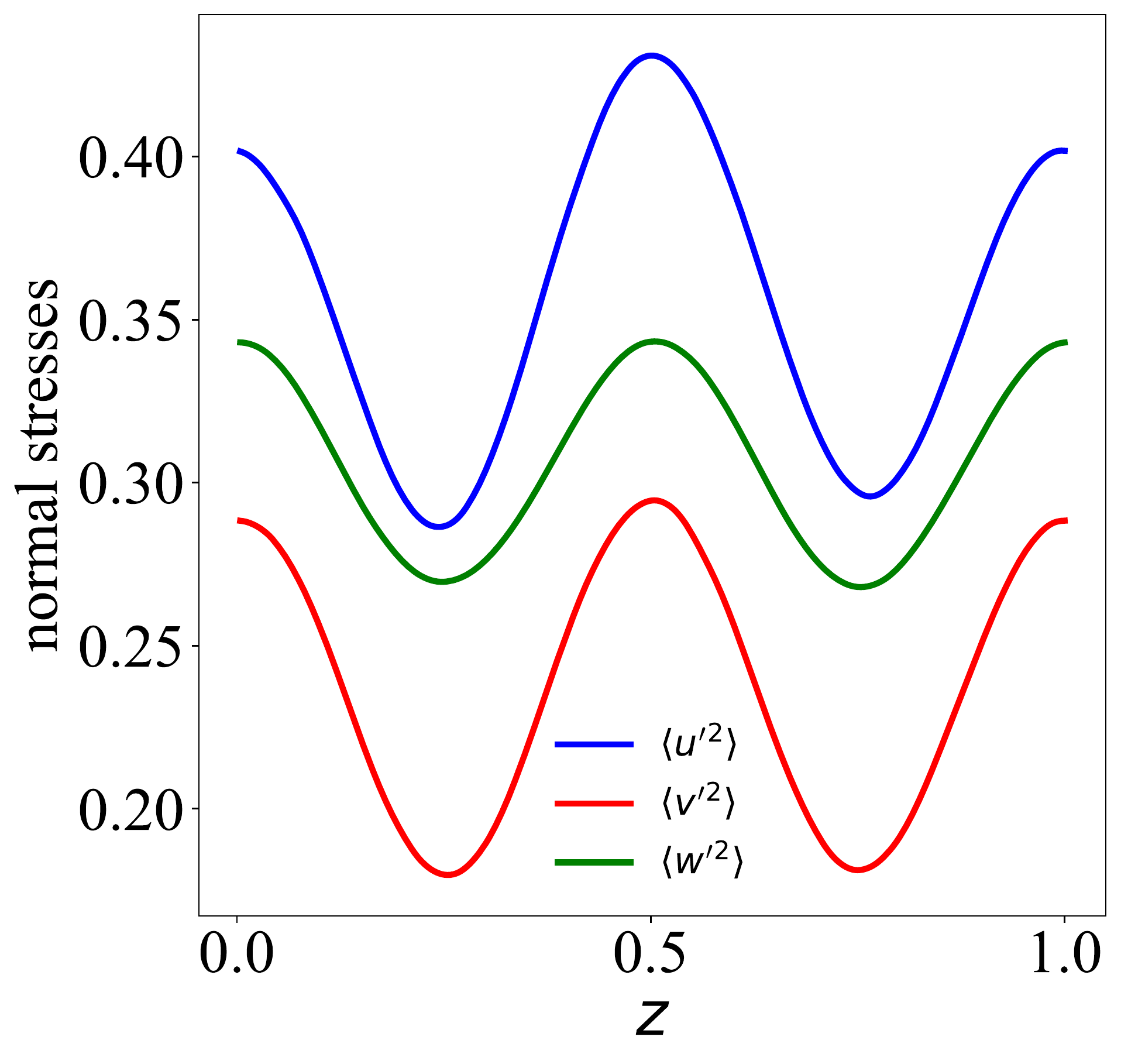}}
}
\caption{The different normal stresses with $\langle u'^2\rangle > \langle w'^2\rangle > \langle v'^2\rangle$. The $z$-dependence is given by the fits of equation (\ref{stressfit}).  }
\label{V_autovariance_along_z}
\end{figure}

\begin{figure}
\includegraphics[width=0.85\linewidth]{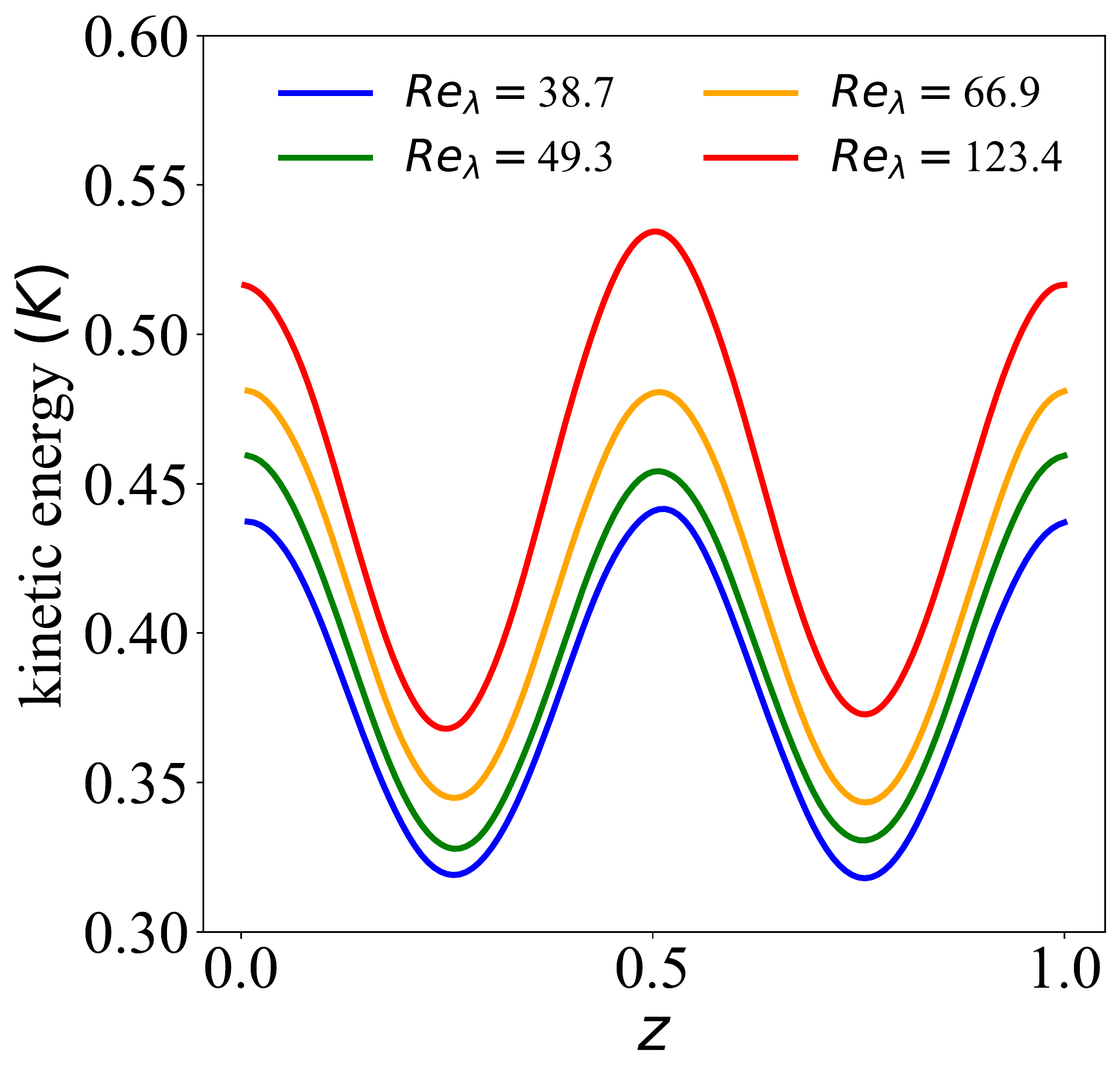}
\caption{The mean kinetic energy profile
$K(z)=\frac{1}{2}\langle u_i u_i \rangle$ for different Reynolds numbers.}
\label{K_along_z}
\end{figure}

For large Reynolds numbers, using equations (\ref{eq_flow_mean})
and (\ref{Uturb}) we find that $\partial_z \tau$ is proportional
to $\sin(2\pi z)$, obtaining finally:
\begin{equation}
\tau =  \frac{1}{2\pi} \left(1 -(2\pi)^2\frac{\kappa}{Re} \right) \cos (2 \pi z).
\label{Stress}
\end{equation}
The first and second moments of the velocity obtained after averaging Navier-Stokes equations are shown in figures \ref{Fig2} a) and b). We observe that only one component of the mean velocity is non-zero; concerning second moments, only the shear stress term $\langle u'w'\rangle$ is non-zero. The turbulence is globally anisotropic since all normal stress components of the stress tensor are different. Specifically, $\langle u'^2\rangle > \langle w'^2\rangle > \langle v'^2\rangle$ (see figure \ref{V_autovariance_along_z}). The diagonal terms have twice the spatial frequency of the forcing. Since $\cos(2 \theta)=2 \cos^2 \theta -1$, they can be written as:
\begin{equation}
 \begin{aligned}
  \langle u'^2\rangle  &= \alpha_1+ \beta_1\cos^2(2\pi z),   \\
   \langle v'^2\rangle  &= \alpha_2+ \beta_2\cos^2(2\pi z),   \\
   \langle w'^2\rangle  &= \alpha_3+ \beta_3\cos^2(2\pi z),
\label{stressfit}
\end{aligned}
\end{equation}
where $(\alpha_i,\beta_i)$ are numerical coefficients expressing the common shape of the three
normal stresses, as visible in figure \ref{V_autovariance_along_z}. The estimated values of such coefficients for the different runs are listed in table \ref{alpha_beta_1_2_3_value}, which will be important for the quadratic closure done in the next section. Consequently, we can write also the evolution of the mean kinetic energy:
\begin{equation}
   K(z)  = \alpha+ \beta\cos^2(2\pi z).
\label{K}
\end{equation}
The mean kinetic energy profiles are presented in figure \ref{K_along_z}, from which we can see that the increase of Reynolds number leads to the global growth of the kinetic energy.
The coefficients $\alpha$ and $\beta$ as functions of $Re_{\lambda}$ are plotted in figure \ref{Kappa_alpha_beta_Vs_Re}(b). The values found for the highest Reynolds number values
are in good agreement  with the values reported in \cite{Borue1996} ($\alpha=0.391$ and $\beta=0.138$)
for simulations of TKF with hyper-viscosity.

\begin{table*}
\caption{\label{alpha_beta_1_2_3_value}The numerical values of the coefficients in Eq. \ref{stressfit} for each run.}
\begin{ruledtabular}
\begin{tabular}{cccccccc }
   No. & $\alpha_1$ & $\beta_1$ & $\alpha_2$ & $\beta_2$ & $\alpha_3$ & $\beta_3$ & $\kappa$ \\
  \hline
   1 & $ 0.2247 \pm 0.0003 $ & $ 0.1121 \pm 0.0004 $ & $ 0.1483 \pm 0.0002 $ & $ 0.0837 \pm 0.0002 $ & $ 0.2279 \pm 0.0004 $ & $ 0.0498 \pm 0.0002 $ & $ 1.0084 \pm 0.0006 $ \\
   2 & $ 0.2291 \pm 0.0003 $ & $ 0.1155 \pm 0.0004 $ & $ 0.1545 \pm 0.0002 $ & $ 0.0885 \pm 0.0002 $ & $ 0.2271 \pm 0.0004 $ & $ 0.0504 \pm 0.0002 $ & $ 1.0253 \pm 0.0006 $ \\
   3 & $ 0.2282 \pm 0.0003 $ & $ 0.1152 \pm 0.0004 $ & $ 0.1622 \pm 0.0002 $ & $ 0.0886 \pm 0.0002 $ & $ 0.2378 \pm 0.0004 $ & $ 0.0553 \pm 0.0002 $ & $ 1.0273 \pm 0.0006 $ \\
   4 & $ 0.2327 \pm 0.0003 $ & $ 0.1141 \pm 0.0004 $ & $ 0.1665 \pm 0.0002 $ & $ 0.0920 \pm 0.0003 $ & $ 0.2483 \pm 0.0004 $ & $ 0.0587 \pm 0.0002 $ & $ 1.0438 \pm 0.0006 $ \\
   5 & $ 0.2385 \pm 0.0001 $ & $ 0.1223 \pm 0.0002 $ & $ 0.1704 \pm 0.0001 $ & $ 0.0951 \pm 0.0001 $ & $ 0.2489 \pm 0.0002 $ & $ 0.0643 \pm 0.0001 $ & $ 1.0734 \pm 0.0004 $ \\
   6 & $ 0.2422 \pm 0.0004 $ & $ 0.1291 \pm 0.0007 $ & $ 0.1749 \pm 0.0003 $ & $ 0.1034 \pm 0.0004 $ & $ 0.2674 \pm 0.0009 $ & $ 0.0772 \pm 0.0004 $ & $ 1.1288 \pm 0.0012 $ \\
   7 & $ 0.2605 \pm 0.0003 $ & $ 0.1394 \pm 0.0005 $ & $ 0.1749 \pm 0.0002 $ & $ 0.1092 \pm 0.0003 $ & $ 0.2678 \pm 0.0006 $ & $ 0.0744 \pm 0.0003 $ & $ 1.1325 \pm 0.0009 $ \\
   8 & $ 0.2766 \pm 0.0019 $ & $ 0.1364 \pm 0.0027 $ & $ 0.1811 \pm 0.0013 $ & $ 0.1092 \pm 0.0016 $ & $ 0.2590 \pm 0.0027 $ & $ 0.0714 \pm 0.0013 $ & $ 1.1242 \pm 0.0049 $ \\
\end{tabular}
\end{ruledtabular}
\end{table*}

\begin{figure}
\centerline{
\includegraphics[width=0.85\linewidth]{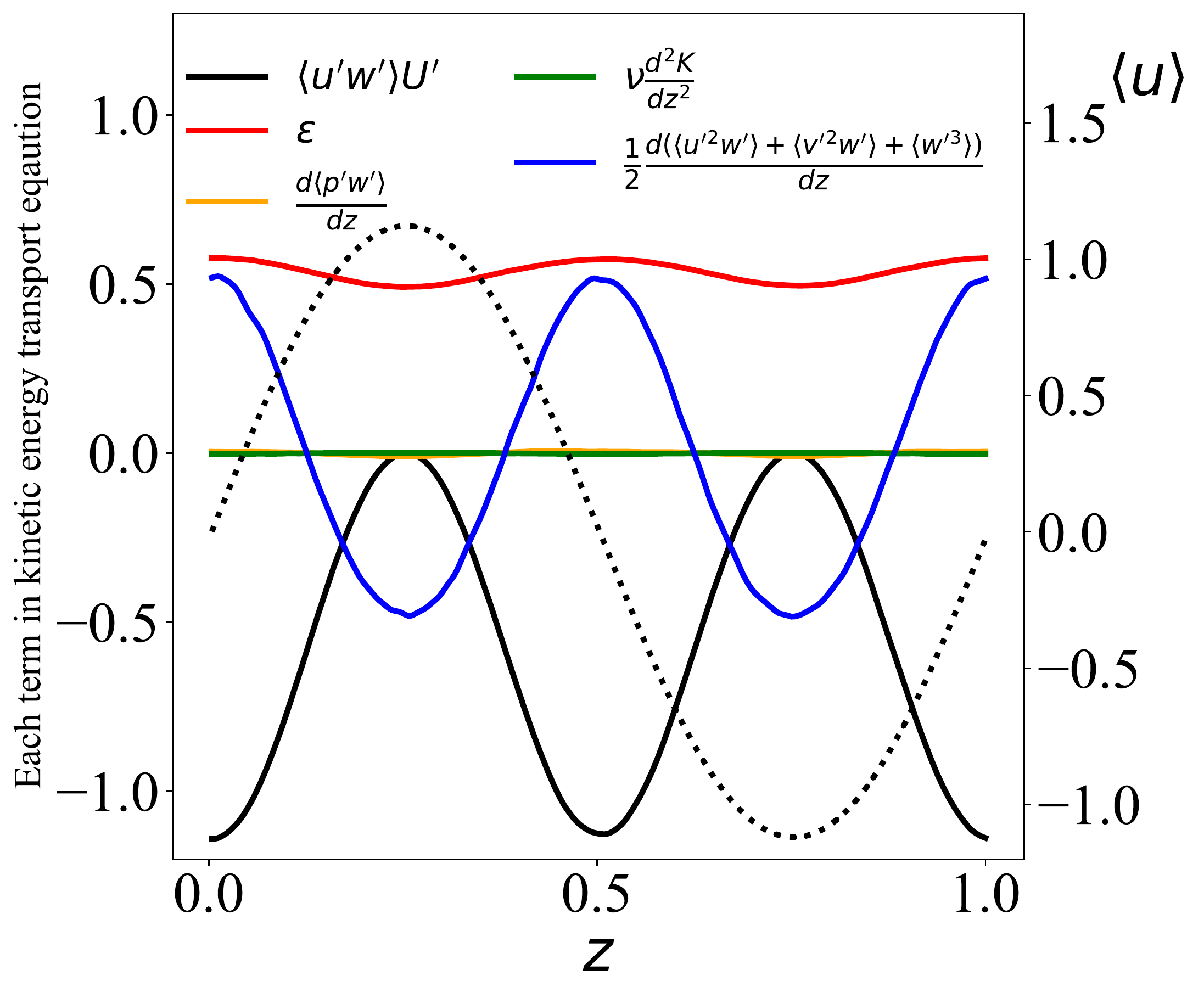}
}
\caption{The amplitudes of the terms of the 
kinetic energy transport equation (\ref{Ktransport}) for Run 7. The mean velocity profile
is also shown as dotted line, for reference.}
\label{terms_of_kinetic_energy_along_z}
\end{figure}

\subsection{Kinetic energy balance equation}
We next consider the kinetic energy balance equation, giving\cite{Bernard2002}: 
\begin{equation}
 \begin{aligned}
  0=&\tau U'(z)-\epsilon-\frac{1}{\rho}<p'w'>+\nu \frac{d^2K}{dz^2}   \\
   -&\frac{d}{dz}\left(<w'u'^2>+<w'v'^2>+<w'^3> \right)
\label{Ktransport}
\end{aligned}
\end{equation}
where the different terms represent respectively the production of kinetic energy,
the dissipation, the pressure work, the viscous transport and turbulent transport.
These terms have been computed for run 7, and are shown in figure \ref{terms_of_kinetic_energy_along_z}. It is visible first that the viscous transport
and pressure works are negligible compared to other terms. There is a balance
between production and dissipation added to turbulent transport. The dissipation is almost
constant, with a small modulation, as already noticed in previous works\cite{Musacchio2014}.
The production is close to 0 at two positions corresponding to vanishing velocity shears. At these positions, the kinetic energy is minimum (see figure \ref{K_along_z}) and
also its transport is negative. The production is larger for strong shear zones, where
 the mean velocity shear is the larger. At these positions, the turbulent transport is
 also the larger.

\section{Expression of the Reynolds stress onto a tensor basis}
In this section we aim at deriving a relation expressing the Reynolds stress in terms of the gradients of the mean velocity flow. Such a relation, also known as turbulence closure equation, allows to have a self-contained model for the description of the mean flow.
To this end, we introduce the Reynolds stress tensor defined as $\mathbf{T}=-\langle \mathbf{u'} \otimes \mathbf{u'}\rangle$ (with $\otimes$ denoting the dyadic product). The anisotropic stress tensor is $\mathbf{R}=-\mathbf{T}+\frac{2}{3}K\mathbf{I}$, where $K$ is the kinematic energy and $\mathbf{I}$ is the identity tensor. The mean velocity gradient tensor $\mathbf{A}=\partial \langle u_i \rangle /\partial x_j$,
and the mean strain-rate $\mathbf{S}$ and rotation-rate $\mathbf{W}$ tensors are also introduced as:
\begin{eqnarray}
     \mathbf{S}&=&\frac{1}{2}\left( \frac{\partial \langle u_i \rangle}{\partial x_j}+\frac{\partial \langle u_j \rangle}{\partial x_i}\right),\\
    \mathbf{W}&=& \mathbf{A}- \mathbf{S}.
\end{eqnarray}
A closure for the turbulence equations corresponds to expressing
the Reynolds stress tensor using mean quantities, e.g. when the closure is local, using
the tensors $\mathbf{S}$ and $\mathbf{W}$.
Below we first consider the simplest linear closure and estimate
the eddy-viscosity, and later on we address a nonlinear expression
using a quadratic constitutive equation.

\subsection{Boussinesq's eddy-viscosity hypothesis and its assessment}
It is seen from equations (\ref{Uturb}) and (\ref{Stress}) that the only non-zero non-diagonal
term in the stress tensor has the same $z$-dependence as the mean gradient term. This leads to
an
eddy-viscosity of the form:
\begin{equation}
 \label{nu_T}
  \nu_T = \frac{\tau}{U'(z)}  =  \left(\frac{1}{(2\pi)^2\kappa}-\frac{1}{Re}  \right).
\end{equation}
The eddy-viscosity does not depend on $z$, but depend on the Reynolds number and the
coefficient $\kappa$. The values of $\nu_T$ provided by this equation
are shown in Table \ref{parameters}; these values are in agreement with results at comparable Reynolds number\cite{Musacchio2014} and imply that $\nu_T/\nu = O(10^2)$.
However, the estimation of an eddy-viscosity does not validate the
linear closure. The Boussinesq's hypothesis, which is at the basis of all
eddy-viscosity turbulence models, corresponds to a linear proportionality between tensors \cite{Boussinesq1877} :
\begin{equation}
  \mathbf{R} = 2\nu_T \mathbf{S}.
\end{equation}
For the flow considered here, there are some symmetries so that the strain as
well as the stress have a simplified form:
  \begin{equation}
\mathbf{S} =  \frac{a}{2} \left(
     \begin{array}{ccc}
       0         &       0     &      1  \\
       0         &       0     &      0  \\
       1         &       0     &      0
        \end{array}  \right)
\label{eq13}
\end{equation}
and
 \begin{equation}
\mathbf{R} = \left(
     \begin{array}{ccc}
       \frac{2}{3} K -\sigma^2_u   &   0     &      \tau  \\
     0 \! &      \frac{2}{3} K -\sigma^2_v    &     0  \\
                     \tau                &                 0       &   \frac{2}{3} K -\sigma^2_w
        \end{array} \! \right),
\label{eq15}
\end{equation}
where $a=U'(z)$, $\sigma^2_u=\langle u'^2\rangle$ and the same for $\sigma^2_v$ and $\sigma^2_w$.

\begin{figure}
\centerline{
\subfigure[$Re_{\lambda}=38.7$]{\includegraphics[width=0.45\linewidth]{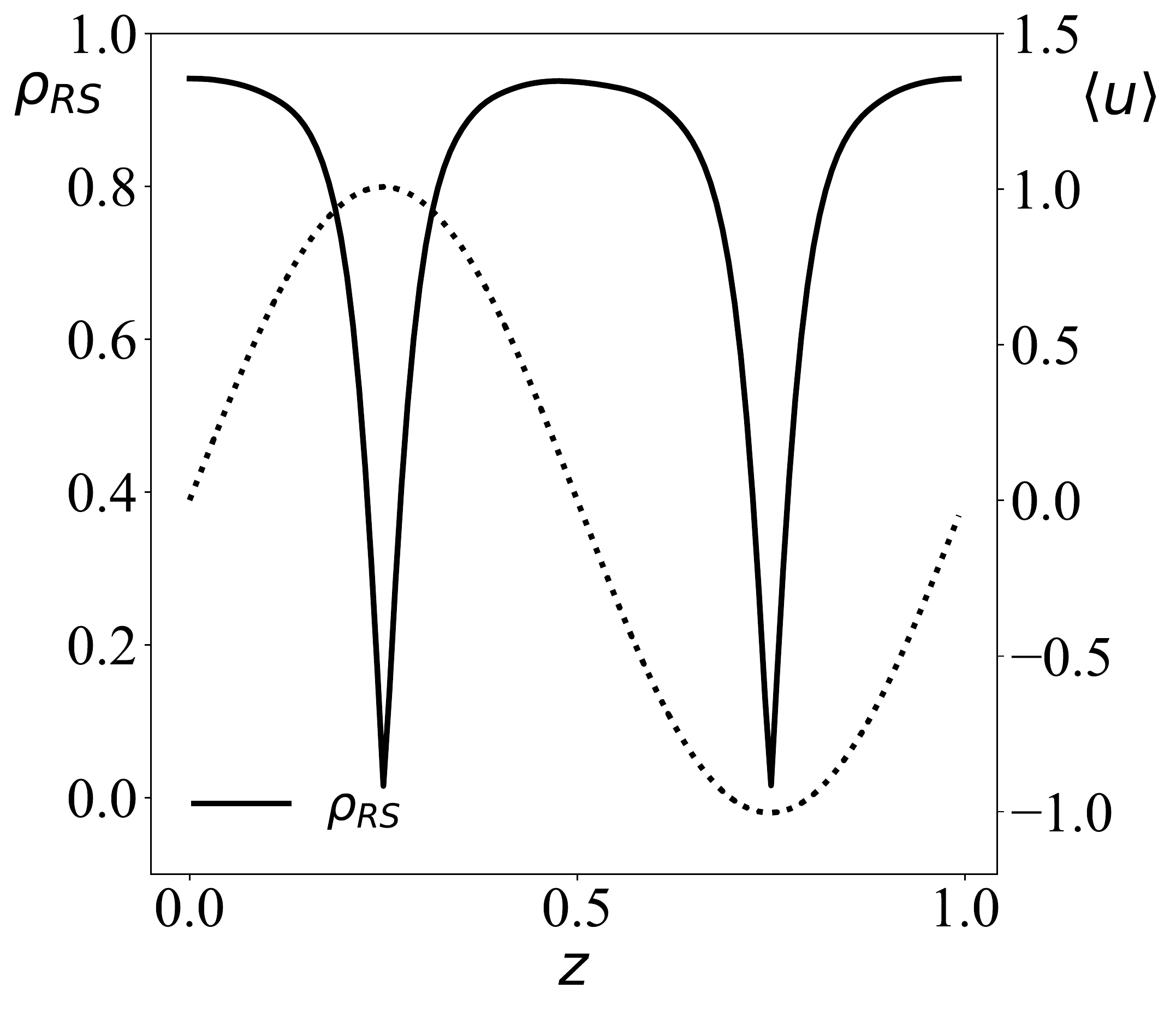}}
\subfigure[$Re_{\lambda}=49.3$]{\includegraphics[width=0.45\linewidth]{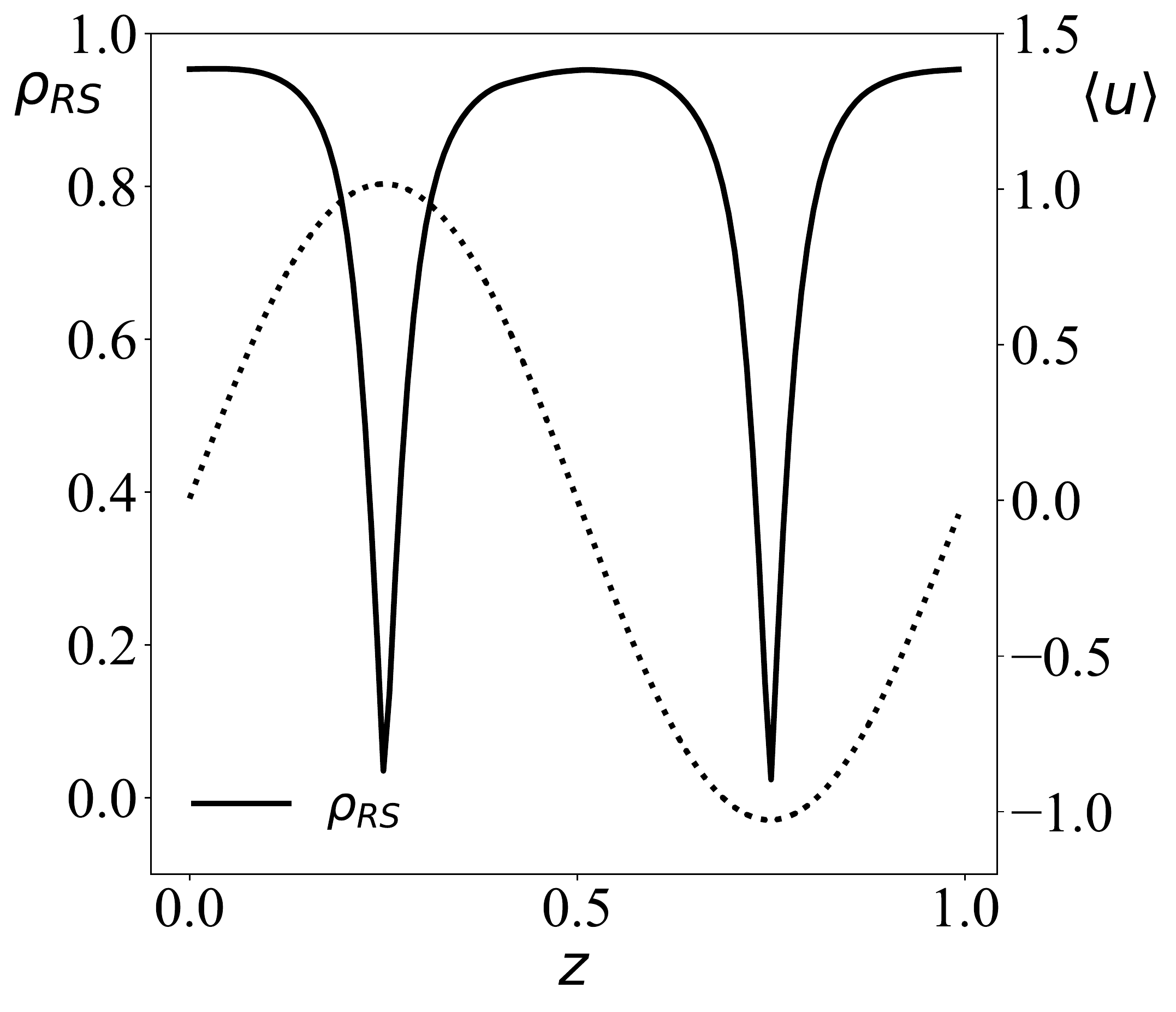}}
}
\centerline{
\subfigure[$Re_{\lambda}=66.9$]{\includegraphics[width=0.45\linewidth]{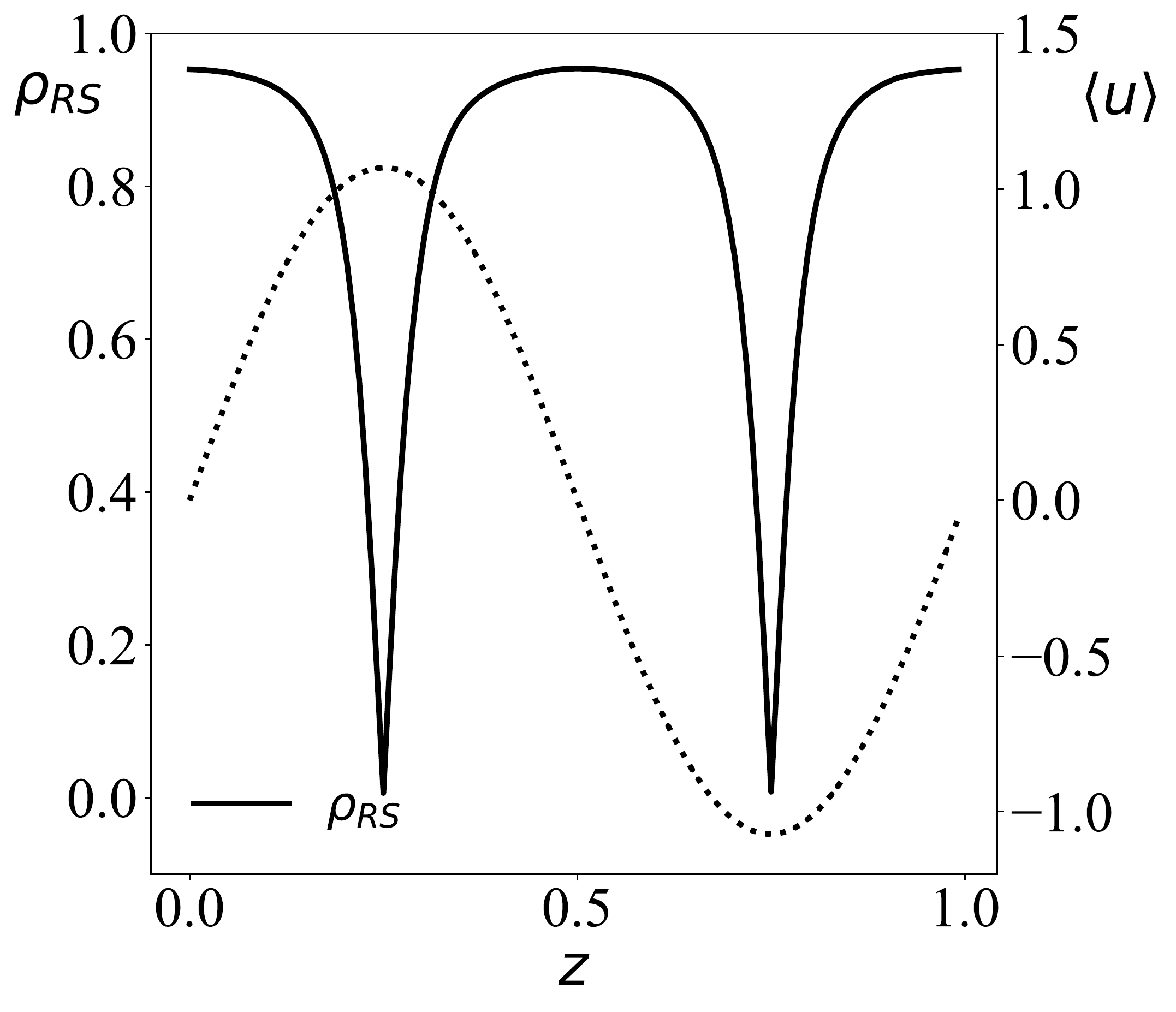}}
\subfigure[$Re_{\lambda}=123.4$]{\includegraphics[width=0.45\linewidth]{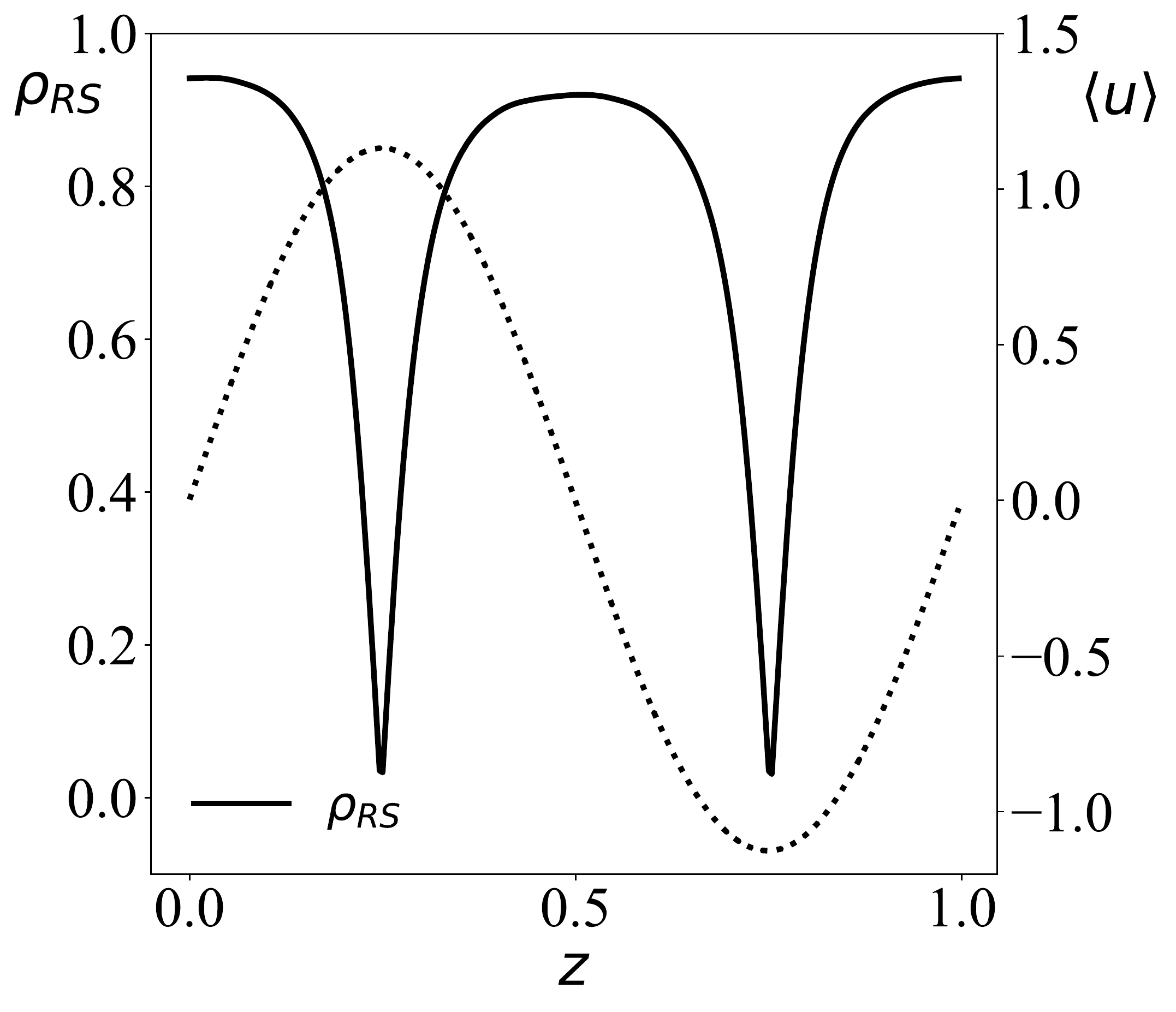}}
}
\caption{Simulation results for the  test of the validity of Boussinesq's hypothesis, representing the alignment $\rho_{RS}$ between $\mathbf{R}$ and $\mathbf{S}$. The mean velocity profile is superposed in dotted line for reference.}
\label{rho_RS}
\end{figure}

It is then clear, as also the case for
turbulent channel flows \cite{Speziale1987,Nisizima1987,Pope2000}, that such linear relation between tensors can
be realized only when diagonal terms are zero, i.e. in an isotropic situation. However,
the Kolmogorov flow is anisotropic and as seen
in figure \ref{V_autovariance_along_z}, the three normal stresses are all different, which means that
a precise proportionality does not exist. In such
framework, eddy-viscosity models will only properly capture the shear stress
component, and cannot represent the normal stresses. The relative importance of
these different components is considered below by using an alignment indicator.
For this, we consider the inner
product between tensors: $\mathbf{A}:\mathbf{B}=\{\mathbf{A}^t\mathbf{B}\}=A_{ij}B_{ij}$, where
$\{\mathbf{X} \}$ is a notation for the trace of $\mathbf{X} $. The norm is then $||\mathbf{A}||^2=\mathbf{A}:\mathbf{A}$.
As a direct test of Boussinesq's hypothesis, we first represent here
the normalized inner product of ${\bf R}$ and ${\bf S}$ tensors (which is similar to
the cosine of an ``angle'' between vectors, see \cite{Schmitt2000,Schmitt2007a}):
\begin{equation}
\rho_{RS} = \frac{\mathbf{R}:\mathbf{S}}{||\mathbf{R}||~||\mathbf{S}||}.
\label{testBouss}
\end{equation}
The ratio   $\rho_{RS}$ is thus a number between $-1$ and $1$,
which characterizes the validity of Boussinesq's hypothesis: it is $1$ when
 this hypothesis is valid, and when close to $0$ it corresponds to  the case of ``orthogonal'' tensors. The behaviour of this quantity is shown in figure \ref{rho_RS}.  It is seen that a plateau close to the value one  appears in certain regions; in particular the Boussinesq's hypothesis is approximately valid when the mean velocity gradient is large,  whereas it fails dramatically for some range of values around the positions where the mean velocity gradient vanishes.

More quantitatively, from Run 7, we find $\rho_{RS}=0.93$ for $z=1/2$ and by
choosing a threshold value at $\rho_{RS}=0.9$, we find that $0.9 \le \rho_{RS} \le 1$
for  $z \in [0,0.13] \cup [0.39,0.59] \cup [0.87,1]$. Hence for about half of
the volume (46$\%$) the linear relation between strain and stress tensor
is approximately valid with $\rho_{RS}$ larger than 0.9, while for
the rest of the flow such linear relation fails to a large extent.

Furthermore, by considering figure \ref{terms_of_kinetic_energy_along_z}
providing the different energy transport terms, we see that Boussinesq's hypothesis is 
closest to validity at positions where the turbulent production is larger, and
is totally failing  at positions where there is almost no production and a negative
turbulent transport term, meaning that the local dissipation is the result
of a transport of kinetic energy. 


\subsection{A quadratic development for the Reynolds stress}
We have seen above that the linear closure model
cannot produce an anisotropic Reynolds stress tensor for anisotropic flows such as the Kolmogorov flow.
Pope \cite{Pope1975} has proposed to use the invariant theory in turbulence modeling,
to represent the stress tensor as a development into a tensor basis
composed of symmetric and traceless tensors expressed
as polynomial based on the mean strain and rotation tensors.
Originally it was on the form ${\bf R} = \sum_{i=1}^{10}a_i {\bf T}_i$ with
 $10$ basis tensors.
By considering  a quadratic development,  only three
tensors are used, which is complete for two dimensional flows \cite{Pope1975}, and is also
a good approximation for fully 3-dimensional flows \cite{Jongen1998}.
As it was used for channel flows \cite{Schmitt2007b,Modesti2020}
and for a tube bundle \cite{Wellinger2021}, we propose here to use it also for the TKF.

In this framework, the anisotropic stress tensor writes as
a three-terms development which is also a nonlinear constitutive equation:
\begin{equation}
{\bf R} = a_1 {\bf T}_1+ a_2 {\bf T}_2 + a_3 {\bf T}_3,
\label{constitutive}
\end{equation}
where the three tensors of the basis are all symmetric and traceless \cite{Pope1975}:
\begin{equation}
 \begin{array}{ll}
 {\bf T}_1 =  {\bf S}, &  {\bf T}_2 =  {\bf S W}-{\bf W S}, \\
{\bf T}_3 =  {\bf S}^2-\frac{1}{3}\eta_1 {\bf I} \;\; .&
\end{array}
\end{equation}
The coefficients $a_i$ can be written using scalar invariants of the flow,
which correspond to scalar fields whose values are independent
of the system of
reference. Invariants can be defined as the traces of different tensor products
\cite{Spencer1971}. Some of the first invariants are the following:  $\eta_1  =  \{{\bf S}^2\}$,
$\eta_2  =  \{{\bf W}^2 \}$, $\eta_3  =  \{{\bf S}^3 \}$,
$\eta_4  =  \{ {\bf SW}^2 \}$,$\eta_5  =  \{ {\bf S}^2{\bf W}^2 \}$,
$ \mu_1  =  \{{\bf R}^2\}$,    $\mu_2  =  \{{\bf RS} \}$,
 $\mu_3  =  \{{\bf RSW} \}$, and $  \mu_4  =  \{ {\bf RS}^2 \}$.
All these invariants can be here estimated numerically. The
coefficients $a_1$, $a_2$ and $a_3$ can be expressed using the above invariants by projecting the constitutive
equation (equation (\ref{constitutive})) onto the tensor basis: successive inner
products of this equation with tensors ${\bf T}_i$ provides
a system of scalar equations involving the invariants \cite{Jongen1998}.
For two-dimensional mean flows such as the KF, we have $ \eta_3 =0$
and $\eta_5=\eta_1\eta_2/2$, and the system of scalar equations
is inverted to provide finally the quadratic constitutive equation using
invariants:
\begin{equation}
{\bf R} =\frac{\mu_2}{\eta_1}{\bf S}-
\frac{\mu_3}{\eta_1\eta_2}{\bf T}_2+6\frac{\mu_4}{\eta_1^2}{\bf T}_3\label{nonlinear2}
\end{equation}
where the invariants write for the TKF:
   \begin{gather}
  \eta_1  =  \frac{a^2}{2}; \quad \eta_2= - \frac{a^2}{2}; \quad \mu_2=a \tau
  \label{eq16} \\
  \mu_3  =  \frac{a^2}{4} \left( \sigma^2_u -\sigma^2_w \right),
  \label{eq17}\\
  \mu_4 =  \frac{a^2}{4} \left( \sigma^2_v-\frac{2}{3}K \right).
   \label{inv}
   \end{gather}
The two remaining tensors of the tensor basis are:
 \begin{equation}
\mathbf{T}_2 =  \frac{a^2}{2}\left(
     \begin{array}{ccc}
       -1         &       0     &      0  \\
       0         &       0     &      0  \\
       0         &       0     &      1
        \end{array}  \right)
\end{equation}
and
\begin{equation}
\mathbf{T}_3 =  \frac{a^2}{12}\left(
     \begin{array}{ccc}
       1         &       0     &      0  \\
       0         &       -2     &      0  \\
       0         &       0     &      1
        \end{array}  \right).
\label{matrices}
\end{equation}
The quadratic constitutive equation finally writes, replacing invariants in equation (\ref{nonlinear2}):

\begin{equation}
{\bf R} = \frac{2\tau}{a} {\bf S}
     + \left( \sigma^2_u- \sigma^2_w \right) \frac{1}{a^2}  {\bf T}_2
     + \left(6\sigma^2_v-4K\right) \frac{1}{a^2} {\bf T}_3.
\label{quadratic}
\end{equation}
Equation (\ref{quadratic}) is a quadratic constitutive equation which expresses a nonlinear closure
of the turbulent Kolmogorov flow; the first constant coefficient is twice the eddy-viscosity ($\tau/a = \nu_T$), whereas
the other coefficients are space-dependent.
Such expression belongs to nonlinear-eddy viscosity models (NEVM) \cite{Pope1975,Speziale1987,Nisizima1987}; it is also related
with another well-known family of models called
explicit algebraic Reynolds-stress models (EARSM), which are based on slightly different assumptions \cite{Gatski1993,Xu1996,Girimaji1996,Wallin2000}.


Equation (\ref{quadratic}) can also be seen as a mathematically
simple relation, obtained from a projection onto a three tensor basis;
however, even if mathematically simple, it provides new and interesting information
on the relative importance of the different terms of
this tensorial development according to the position considered. In NEVM and EARSM,
the coefficients of such nonlinear development are expressed using other quantities
such as e.g. $K$ and $\epsilon$, which are computed in the domain considered using
transport equations \cite{Gatski2002}. Here we are not building such a model
but the assessment of the relative importance of each term in the development will
be useful for modelling studies.

\begin{figure}
\centerline{
\subfigure[$Re_{\lambda}=38.7$]{\includegraphics[width=0.45\linewidth]{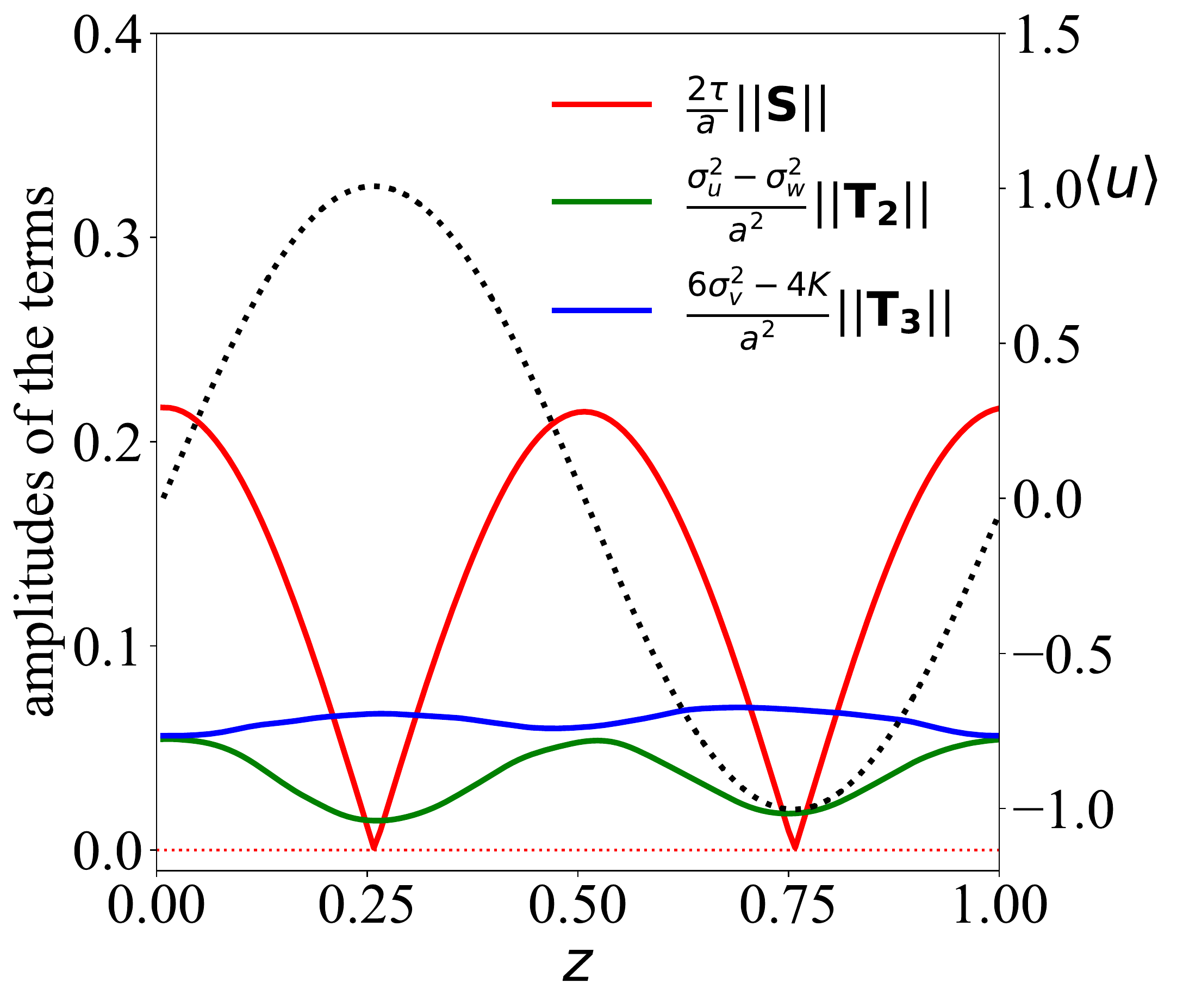}}
\subfigure[$Re_{\lambda}=49.3$]{\includegraphics[width=0.45\linewidth]{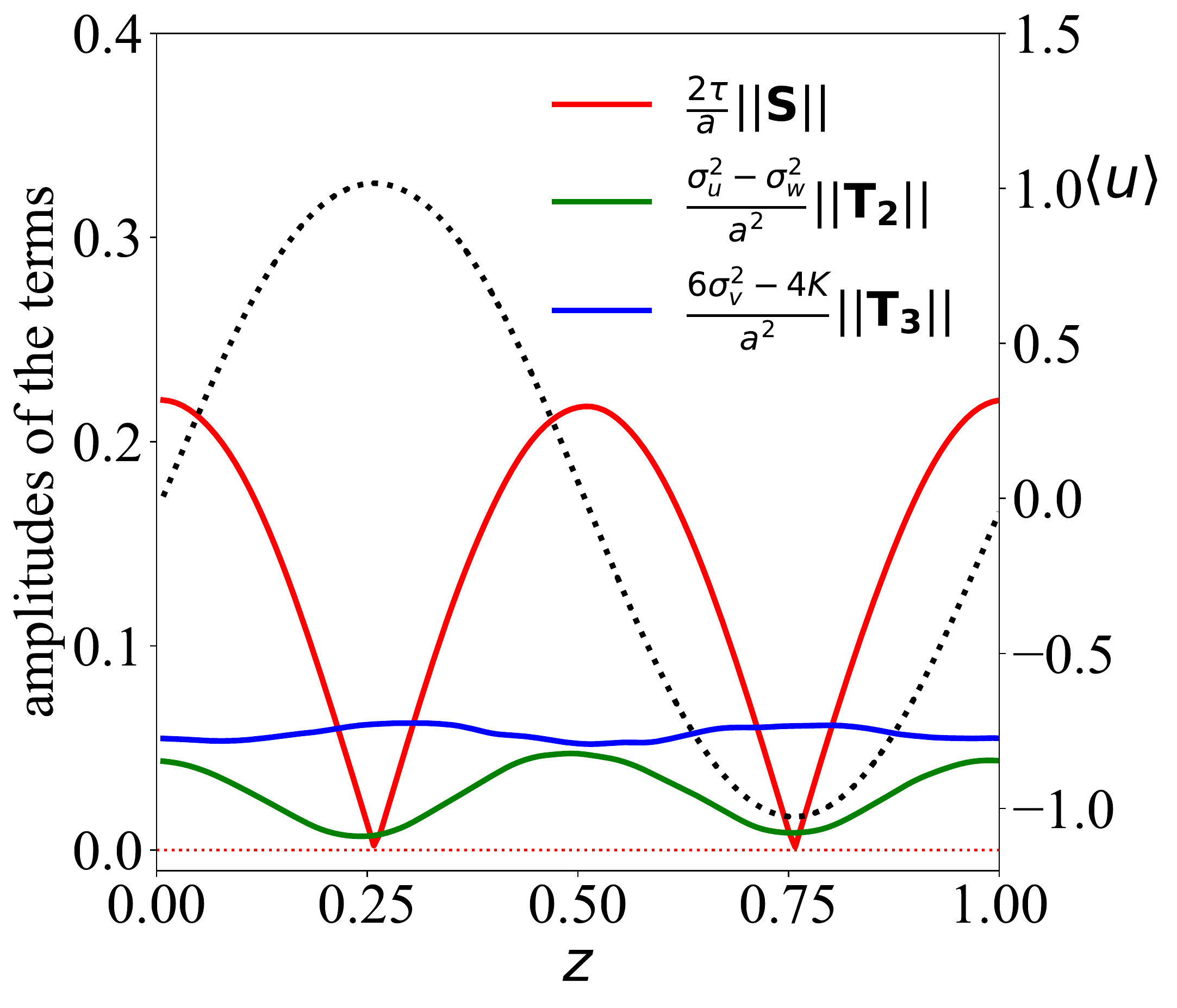}}
}
\centerline{
\subfigure[$Re_{\lambda}=66.9$]{\includegraphics[width=0.45\linewidth]{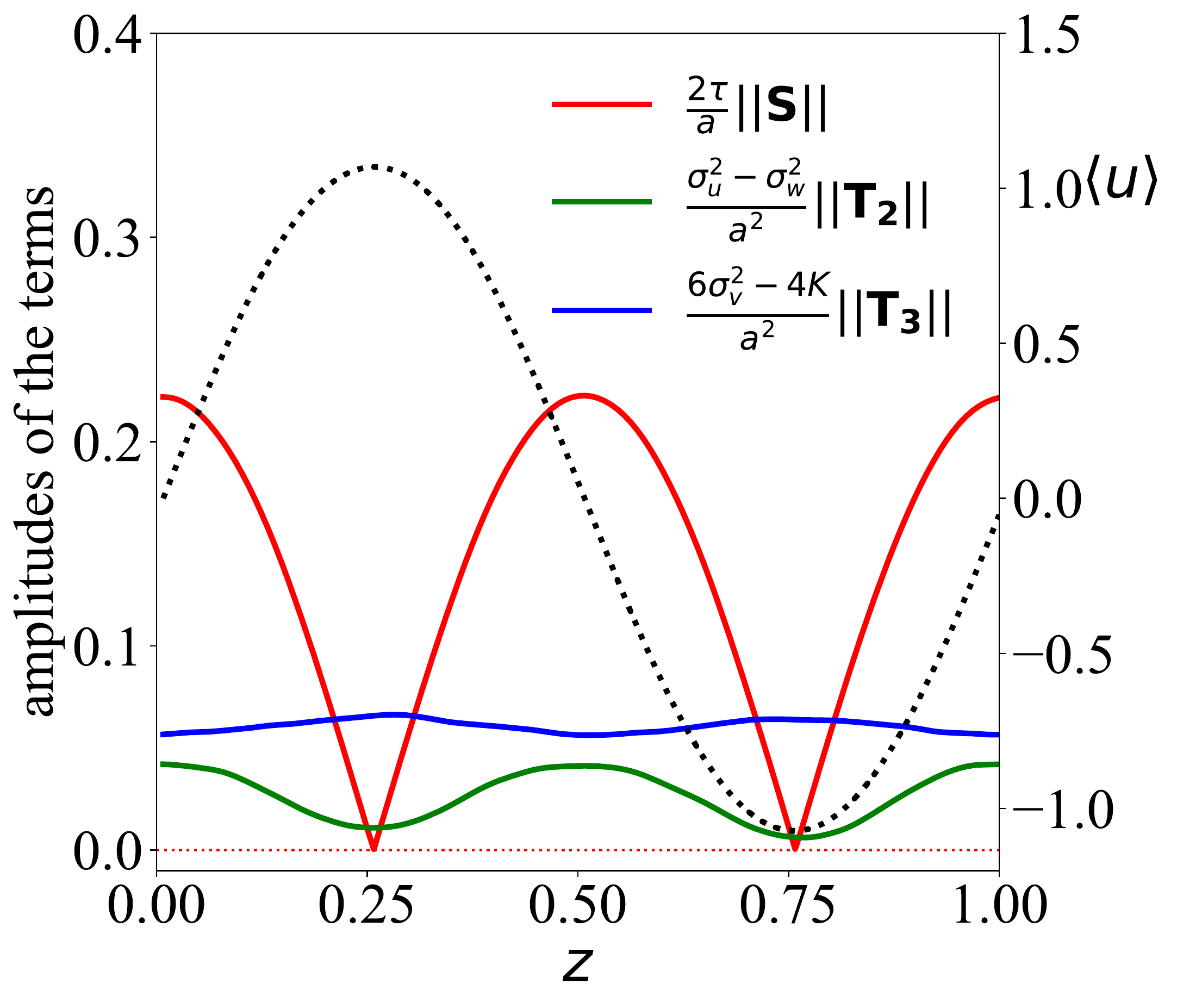}}
\subfigure[$Re_{\lambda}=123.4$]{\includegraphics[width=0.45\linewidth]{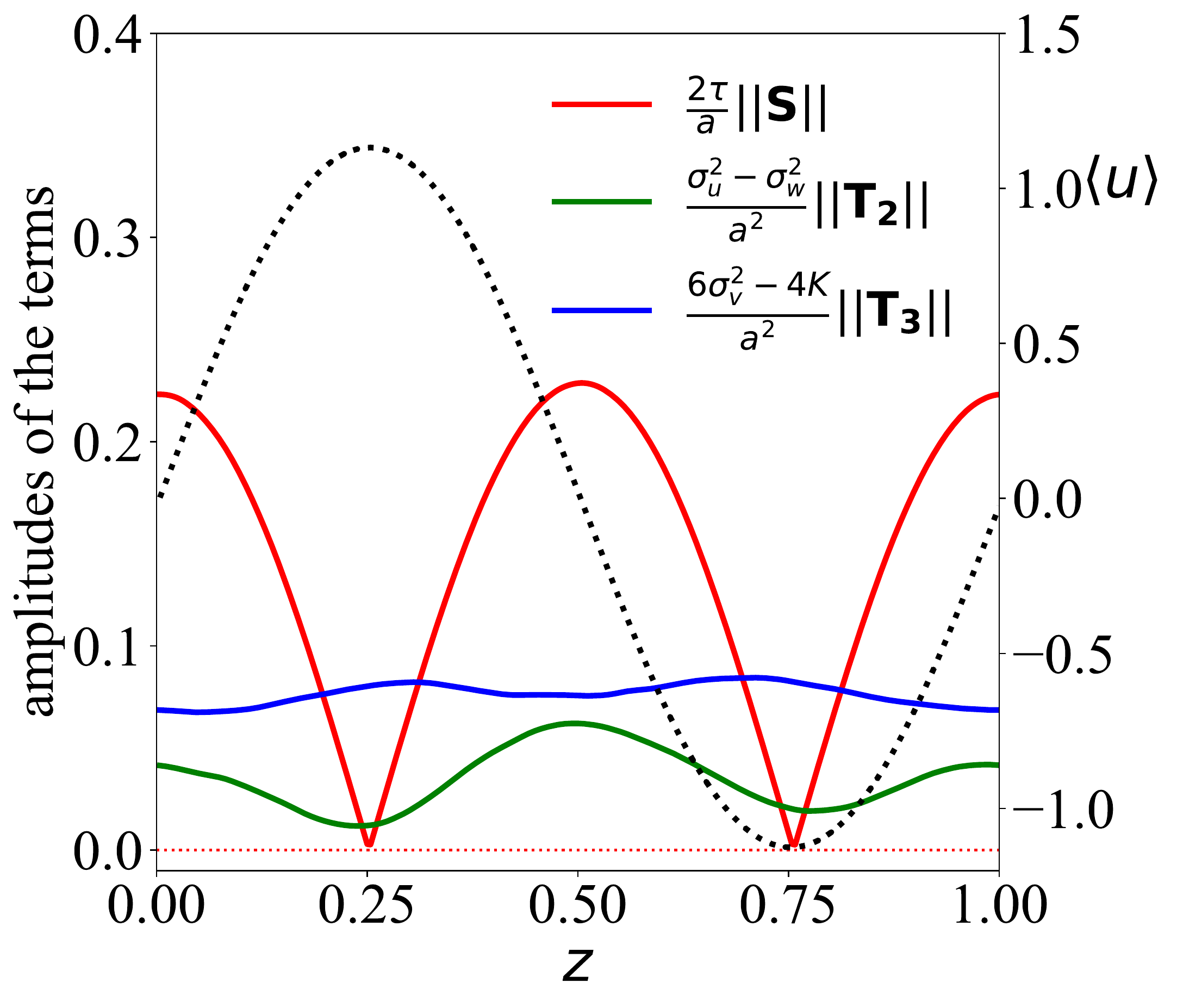}}
}
\caption{The amplitudes of the terms at the right hand side of Eq.~\eqref{quadratic} as function of $z$. The mean velocity profile is also represented as a dotted line, for reference. The horizontal red dotted lines mark the 0 value for the amplitudes. The simulation results of Run 1, 3, 5, 7 are shown here.
}

\label{terms_of_R}
\end{figure}

When $a=U'(z) \simeq 0$, for $z\simeq 1/4$ and $z\simeq 3/4$, $\cos (2 \pi z)=0$
and all ${\bf S}$,  ${\bf T}_2$ and ${\bf T}_3$  vanish, but in the three-terms
development of ${\bf R}$, the second term and the third are non-zero constants,
since the coefficients diverge (the $a^2$ terms cancel).
In those positions, we see that ${\bf R}$ is a diagonal tensor which is not vanishing: figure \ref{terms_of_R} shows that the second term is
also very small and that
the third term is dominant. This means
that in those positions, the Boussinesq's
linear eddy-viscosity approximation is no longer appropriate and the anisotropic stress tensor is a constant perpendicular to the linear term and approximately
proportional to ${\bf T}_3 =  {\bf S}^2-\frac{1}{3}\eta_1 {\bf I} $.
We have also noted above that at those positions,
the production of kinetic energy is very small and the kinetic energy dissipation 
at those positions is produced elsewhere and transported.

\section{Periodic flow with non-sinusoidal forcing}
As we have seen, the Kolmogorov flow, in its original definition, is sustained by a monochromatic sinusoidal forcing. The resulting mean flow profile is also sinusoidal with the same shape of the force term, both in laminar and in turbulent flow conditions. This peculiar property is of great help in the analysis of the turbulent flow and, as we have seen, it simplifies the formulation of a closure relation. It is therefore of interest to ask what happens when the shape of the force is changed to other periodic or quasi-periodic shapes \cite{Romano2021}.
Here we investigate a forcing having a Gaussian shape. Although this function is non-periodic and has an unbounded support, one can adjust its width in such a way that its value and its derivative becomes sufficiently small at the borders. Furthermore, the Gaussian has the advantage to be easily implemented in spectral space.
Here we test only one given Reynolds number, as this is sufficient to contrast the qualitative differences with the sinusoidal forcing case. The dimensionless viscosity for this case are $0.00063$, the same as the Run 4 with sinusoidal forcing, while the Taylor based Reynolds number is $44.1$.

The  Gaussian-type forcing is of the form of:
\begin{equation}\label{gauss}
    \mathbf{f}=(A \exp\left(-\frac{( z-0.5)^2}{2\ell^2} \right)+C)\mathbf{e}_x,
\end{equation}
where $A$ is the forcing parameter, $C$ is a constant to make $\overline{ \mathbf{f} }=0$ in equation (\ref{NS}) and $\ell=0.1$ (in $H$ units), which is equivalent to the standard deviation, and controls the width of the
Gaussian shape.

\begin{figure}
\begin{center}
\subfigure[]{
\includegraphics[width=0.45\linewidth]{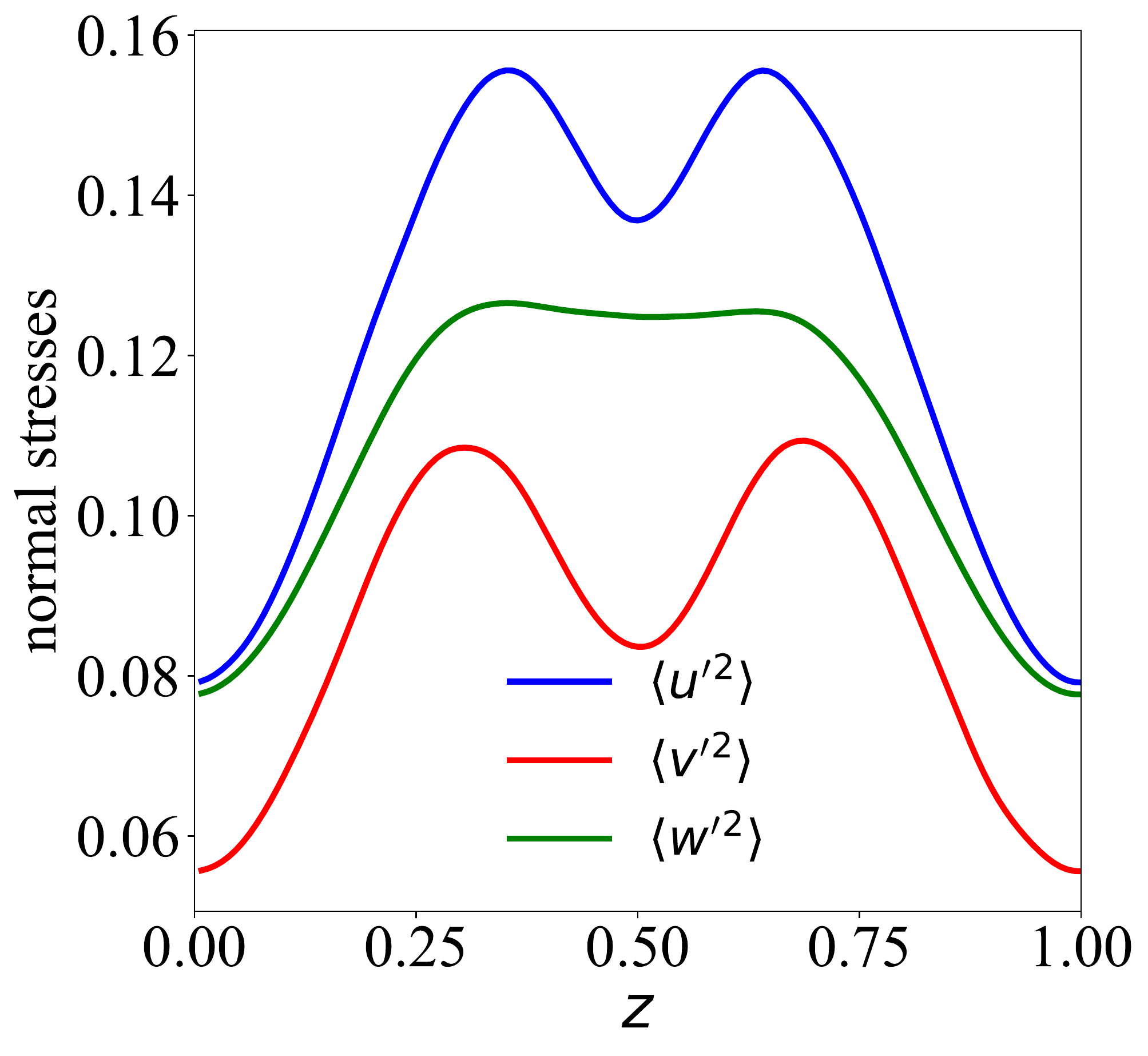}
}
\subfigure[]{
\includegraphics[width=0.47\linewidth]{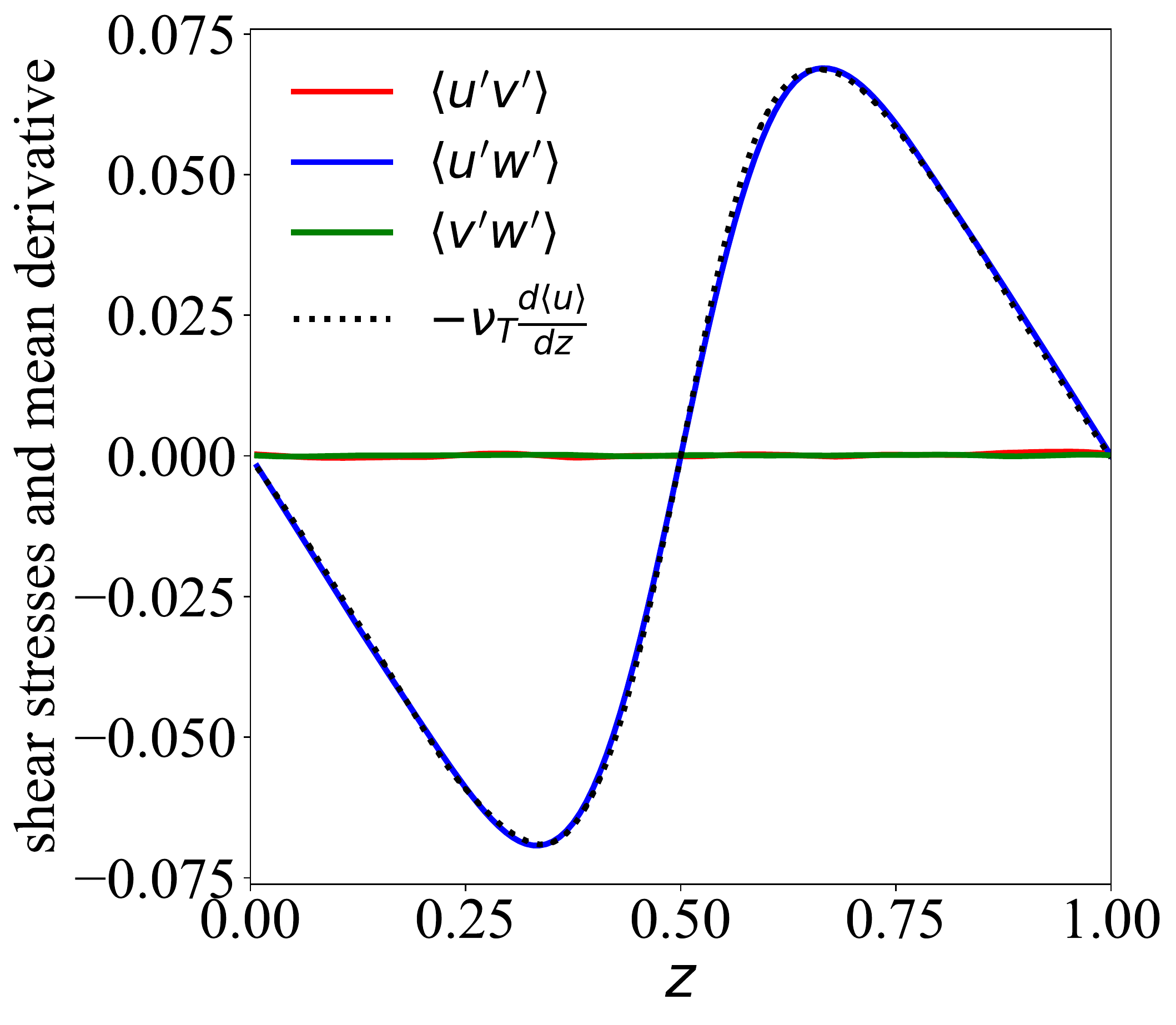}
}
\end{center}
\caption{(a) The normal stresses; (b) The shear stresses and the function of $-\nu_T\frac{d\langle u\rangle}{dz}$ (black dotted line), where $\nu_T$ is the turbulent viscosity (Eq. \eqref{nu_T}) and numerically found as $0.016$ here,  for the case with Gaussian forcing (same parameters with Run 4).}
\label{Gaussian:V_autovariance_along_z_and_V_covariance_along_z}
\end{figure}

Figure \ref{Gaussian:V_autovariance_along_z_and_V_covariance_along_z} shows the Reynolds stress components: the shear stress and normal stresses. The normal stresses have a shape that can not be fitted with known functions and only the numerical result is shown here. It is visible that normal stresses  are again anisotropic, with $\sigma_u^2 > \sigma_w^2 > \sigma_v^2$ at all positions.
The shear stress $\tau = -\langle u'w'\rangle$ is as in the TKF the only non-zero shear stress, and is proportional to $U'(z)$, with a coefficient $\nu_T = 0.016$ (it was found of the same order, $0.0236$, in Run 4 wih sinusoidal forcing according to Eq. \eqref{nu_T}).

\begin{figure}
\begin{center}
\includegraphics[width=0.95\linewidth]{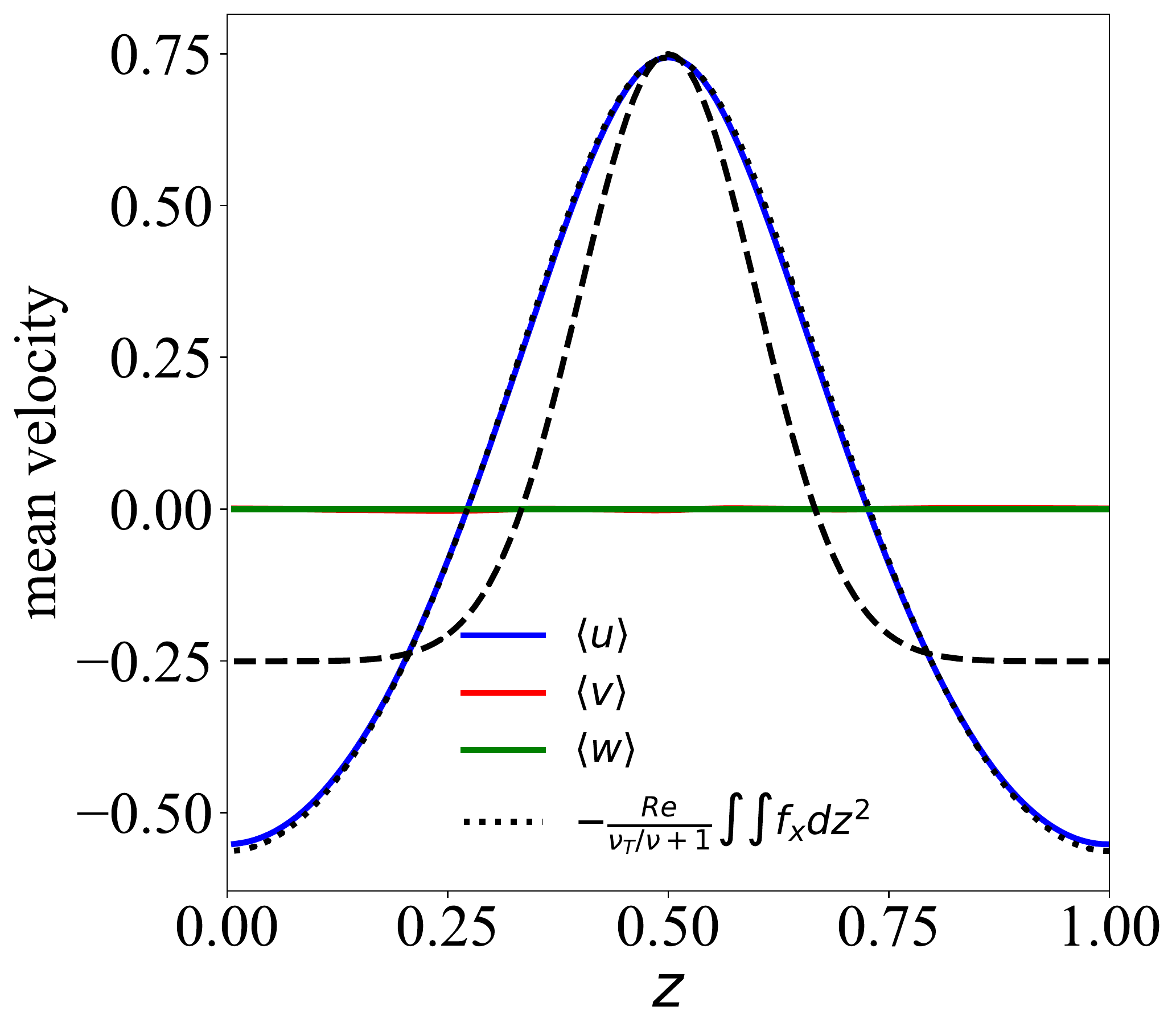}
\end{center}
\caption{Comparison of the mean profile ($\langle u\rangle$,  blue solid line) and the double integral of the forcing (black dotted line) for the case with Gaussian forcing (same parameters with Run 4). The coefficient of $\frac{Re}{\nu_T/\nu+1 }$ is numerically found as 61.27. The black dashed line represents the Gaussian profile forcing (\ref{gauss}). The profiles of $\langle v\rangle$ and $\langle w\rangle$
are very close to zero and almost superposed. }
\label{Gaussian:mean_velocity_along_z}
\end{figure}

This shows that here also the eddy-viscosity is a constant, as was found for the sinusoidal forcing. From equation (\ref{eq_flow_mean}), introducing this numerical result
$\tau = \nu_T U'(z)$ we obtain
\begin{equation}
   U''(z) = -\frac{Re}{\nu_T/\nu+1 } \mathbf{f}.
\label{Usecond}
\end{equation}
This shows that the mean velocity profile is proportional to twice the
integral of the forcing. The primitive of the Gaussian is non-analytical
and involves the error function; it can be estimated numerically, as shown in figure \ref{Gaussian:mean_velocity_along_z}.
An excellent superposition is found. The shape of the mean
velocity is still Gaussian-like, but its precise analytical expression is given by the function whose second derivative is a Gaussian.

\begin{figure}
\centerline{
\includegraphics[width=0.95\linewidth]{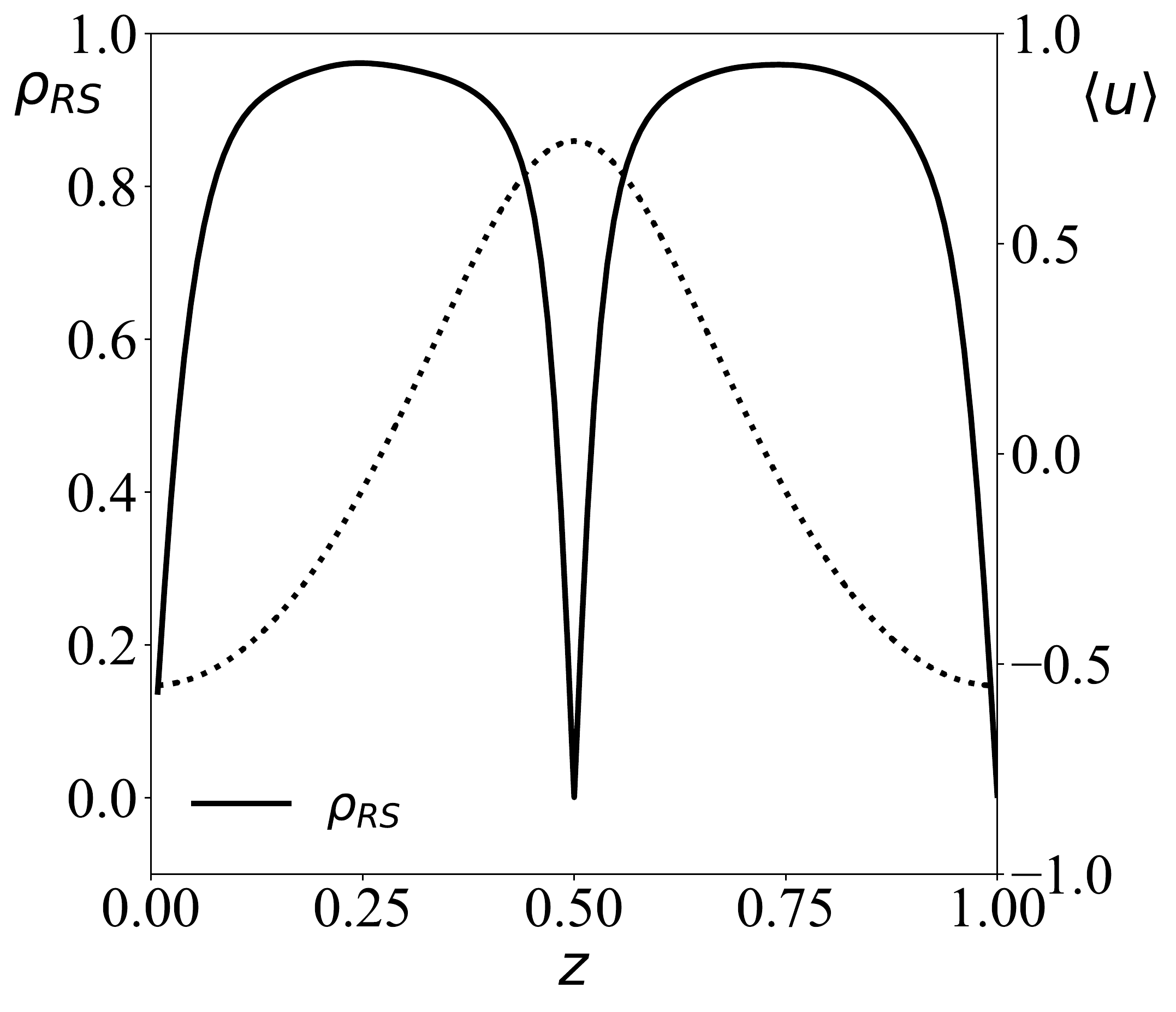}  }
\caption{Estimation of the alignment $\rho_{RS}$ between $\mathbf{R}$ and $\mathbf{S}$, for the case with Gaussian forcing (same parameters with Run 4). The mean velocity profile is superposed in dotted line for reference.}
\label{Gaussian:rho_RS}
\end{figure}

The alignment between the tensors of ${\bf R}$ and ${\bf S}$ (as defined in Eq. \eqref{testBouss}) is also examined for the case with Gaussian forcing, as shown in Fig.  \ref{Gaussian:rho_RS}. We observe also in this case the plateaus  obtained at the position where the mean velocity gradient is large. Qualitatively, we find that $0.9 \le \rho_{RS} \le 1$ for  $z \in [0.12,0.4] \cup [0.6,0.88]$, corresponding totally to about $56\%$ of the considered domain.

\begin{figure}
\begin{center}
\includegraphics[width=0.95\linewidth]{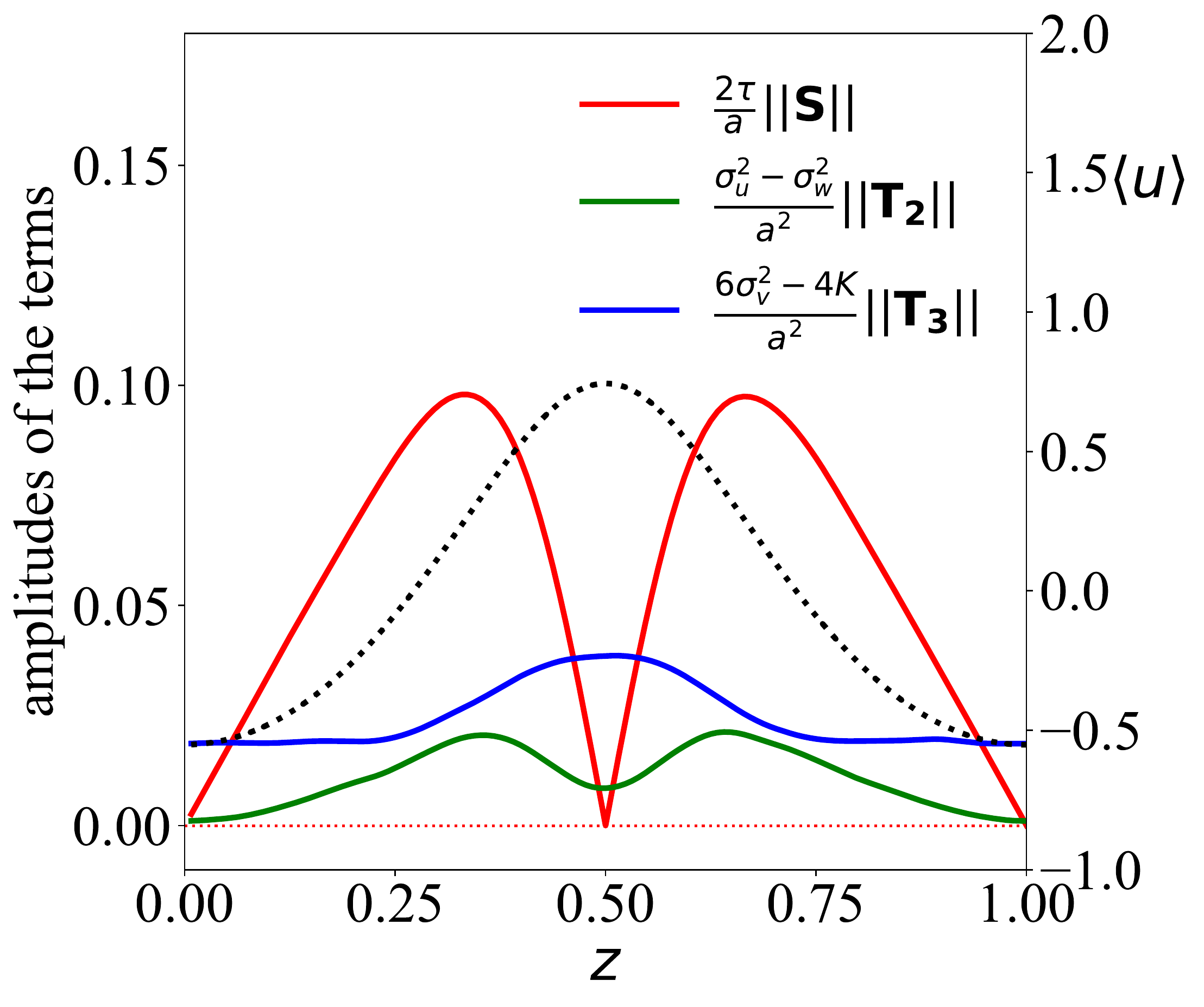}
\end{center}
\caption{The amplitudes of the terms at the right hand side of Eq.~\eqref{quadratic} as function of $z$ for the case with Gaussian forcing (same parameters with Run 4). The mean velocity profile is also represented as a dotted line, for reference. The horizontal red dotted lines mark the 0 value for the amplitudes.
}
\label{Gaussian:terms_of_R}
\end{figure}

Next, the quadratic development given by equation (\ref{quadratic})
is tested and shown in figure \ref{Gaussian:terms_of_R}. It is seen that
the linear term is dominant in part of the domain, and vanishes at the
central position, where the mean velocity is null; in this position the
two nonlinear terms do not vanish. Globally all three terms are needed to
achieve the closure of the stress tensor.




\begin{figure}
\centerline{
\includegraphics[width=0.45\linewidth]{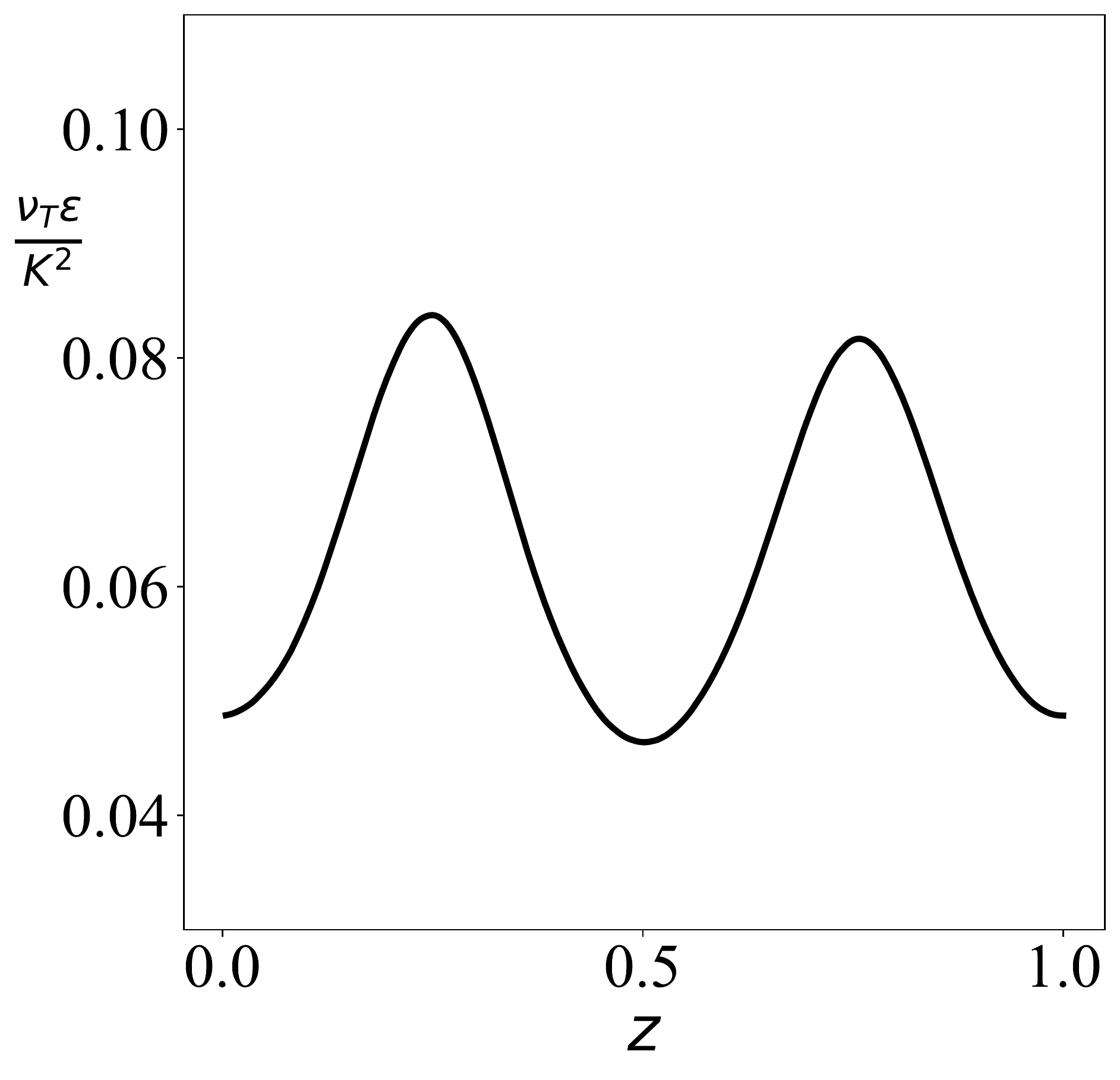}
\includegraphics[width=0.45\linewidth]{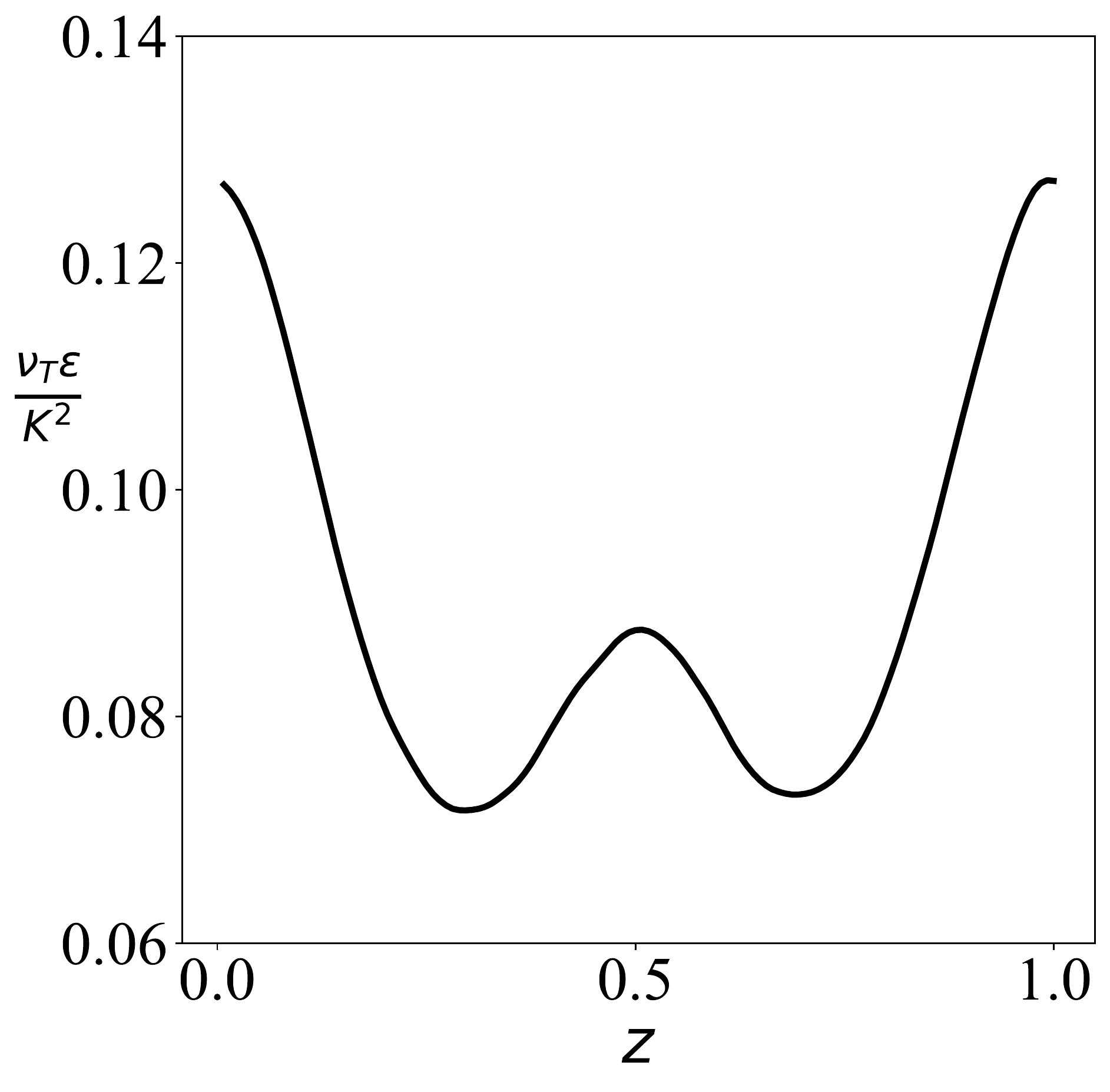}
}
\caption{The adimensional mean quantities of $\frac{\nu_T\epsilon}{K^2}$ of (a) Run 7 and (b) the case with Gaussian forcing.}

\label{FigNew}
\end{figure}

\section{Conclusion}
We have considered here the closure relation for the Reynolds stress in a numerically simulated
turbulent Kolmogorov flow. As the simplest realization of turbulence with a spatially dependent mean flow, such  model system is a convenient test ground for turbulent transport models.
With a forcing of the form $\sin(2\pi z)$, it was found that the mean velocity profile
has the same form, with a damping of a factor $\kappa $
with respect to the mean velocity value
calculated from forcing terms. The value of $\kappa$ was found to increase with the Reynolds number, and of the order of 1.01 to 1.12 for the range of
Reynolds numbers considered here. The only non-zero shear stress term is proportional to
$\cos(2\pi z)$ as expected, and the normal stress components all involve
a square cosine expression of the form $\alpha + \beta \cos^2(2\pi z/H)$,
where the parameters $\alpha$ and $\beta$
are numerically estimated and found to saturate for the largest Reynolds numbers considered here. The normal stresses, i.e. $\langle u'^2\rangle , \langle w'^2\rangle , \langle v'^2\rangle$, are all different in amplitude,
showing that the turbulence is anisotropic.

It was also shown that a quadratic nonlinear constitutive equation
can be proposed for this flow. Specifically a linear term and two
nonlinear terms in the form of traceless and symmetric tensors
${\bf S W}-{\bf W S}$ and ${\bf S}^2-\frac{1}{3}\{{\bf S}^2 \} {\bf I}$ are involved and
their coefficients are here numerically estimated. For about half of the flow domain, the linear
term is dominating, whereas for the vanishing mean velocity regions a constant term is the only one remaining.
Hence an effective viscosity
coefficient can indeed be estimated for the Kolmogorov flow, but contrary to what has been
indicated previously \cite{Rollin2011} this type of turbulence without
boundaries does not generate an effective diffusion of momentum,
since nonlinear terms are needed: globally all linear and nonlinear terms are needed for the complete Reynolds stress closure.

The values obtained here are in agreement with
previous works \cite{Borue1996,Musacchio2014}.
Using
8 different runs with different grid sizes from $128^3$ to $512^3$, and with Reynolds numbers from
$Re_{\lambda}=39$ to $198$, the Reynolds number dependence of involved parameters has been checked with expected convergence toward the largest Reynolds number considered.

Finally a periodic flow with non-sinusoidal forcing has been considered, with the choice of a Gaussian shape.
It was found that the shear stress term $\tau$ is proportional to the mean velocity derivative, indicating that for such forcing also the eddy-viscosity does not depend on $z$. With such numerical result, we obtain that the mean velocity profile is twice the integral of the forcing. The shape of the normal stresses in this case is non-trivial and can not be precisely fitted. A quadratic development of the constitutive equation can also be proposed for this flow.

We may qualitatively compare here the turbulent flows sustained by the sinusoidal forcing (TKF) and
the Gaussian forcing. In both cases the eddy-viscosity is found to be constant. We observe that the
relation stating that $U''$ is proportional to the forcing
(\ref{Usecond}), is in fact also valid for the TKF case. Such
constant eddy-viscosity found here for two very different types of forcing is not
to be taken as a coincidence: we  hypothesize here that this is a general property of
such boundary free periodic flows. We shall note here that for such flows
the classical expression of the eddy-viscosity $\nu_T=C_{\mu}k^2/\epsilon$
does not hold, as demonstrated in Figure \ref{FigNew} plotting $\frac{\nu_T\epsilon}{K^2}$ for Run 7 and Gaussian forcing cases, showing that $C_{\mu}$ is not a constant for both flows.
There are however, two main differences between the two forcing cases. The first lies
in the shape of the mean velocity profile. For the TKF,
since the second integral of the forcing is proportional to the forcing, equation
 (\ref{Usecond}) directly gives the mean velocity profile, as being proportional to the forcing. In the Gaussian forcing case,
  the mean velocity is a non-analytical function, obtained as the second integral of the
  Gaussian. The second difference is in the shape of normal stresses, whose expression could be fitted using $\cos^2$ terms for the TKF, while no known analytical fit for the Gaussian forcing is available.


It is also worth mentioning that the linear or non-linear eddy-viscosity modelling that were considered in this work all rely on a local expression of the velocity,
through derivatives of the mean velocity field. Such local expression
is known to be incomplete \cite{Tennekes1972,Corrsin1975,Pope2000} and non-local models
have been proposed, based on space and time integrations of velocity gradients \cite{Hinze1974,Hamba2005}, as reviewed and discussed in a recent book \cite{egolf2020}.

As a perspective let us mention the recent work \cite{Bos2020} describing
a flow behind a grid in a wind tunnel as having locally, close to the grid,
sinusoidal variations. This could provide ideas to perform measurements of
a TKF and to check experimentally the closure proposed in the present study.
A more systematic numerical study of non-sinusoidal forcing in future work may also help to provide a general expression for normal stresses, which would be valid for all kind of forcing.
It remains also
to be understood from analytical arguments why the eddy-viscosity does
not depend on $z$ for such periodic flow, contrary to what is found in similar bounded flow such as the channel flow \cite{Schmitt2007b} or the boundary-layer flow.


%
%

%

\begin{acknowledgments}
The comments of two reviewers that helped to improve
 this paper are acknowledged. This work is under the joint support of Shanghai Jiao Tong University and the
French Region ``Hauts-de-France'' in the framework of a cotutella PhD programme.
We thank Dr. Michael Gauding (CORIA (CNRS UMR 6614), Rouen, France) for providing a numerical code
which was adapted for our specific present topic.
We acknowledge the computing resources including the High Performance Computing Center
(HPCC) at Universit\'e de Lille, CALCULCO of Universit\'e
 du Littoral C\^ote d'Opale and the National Supercomputer Center in Guangzhou, China.
\end{acknowledgments}

\section*{Data Availability Statement}
Data available on request from the authors.

\section*{Conflict of interest}
The authors have no conflicts to disclose.

%

\end{document}